\documentclass[eat,twocolumn]{jmlr}
   
\usepackage{longtable}

\usepackage{booktabs}
\usepackage[load-configurations=version-1]{siunitx} 

\newcommand{\cs}[1]{\texttt{\char`\\#1}}

\theorembodyfont{\upshape}
\theoremheaderfont{\scshape}
\theorempostheader{:}
\theoremsep{\newline}
\newtheorem*{note}{Note}

\jmlrvolume{1}
\firstpageno{1}

\jmlryear{2022}
\jmlrworkshop{Machine Learning for Health (ML4H) 2022}

\title[HLARIMNT]{ Efficient HLA imputation from sequential SNPs data by Transformer}

   \author{\Name{Kaho Tanaka}${}^\text{ 1,2}$\Email{tanaka.kaho.58x@st.kyoto-u.ac.jp}\\
    \Name{Kosuke Kato}${}^\text{ 2}$\Email{kosuke.kato@riken.jp}\\
   \Name{Naoki Nonaka}${}^\text{ 2}$ \Email{naoki.nonaka@riken.jp}\\
   \Name{Jun Seita}${}^\text{ 2}$ \Email{jun.seita@riken.jp}\\
   \addr 1. Faculty of Engineering, Kyoto University \\
   \addr 2. Advanced Data Science Project, RIKEN Information R\& D and Strategy Headquarters
   }
   
\begin{document}

\maketitle

\begin{abstract}
 Human leukocyte antigen (HLA) genes are associated with a variety of diseases, however direct typing of HLA is time and cost consuming. Thus various imputation methods using sequential SNPs data have been proposed based on statistical or deep learning models, e.g. CNN-based model, named DEEP*HLA. However, imputation efficiency is not sufficient for infrequent alleles and a large size of reference panel is required. Here, we developed a Transformer-based model to impute HLA alleles, named "HLA Reliable IMputatioN by Transformer (HLARIMNT)" to take advantage of sequential nature of SNPs data. We validated the performance of HLARIMNT using two different reference panels; Pan-Asian reference panel ($n=530$) and Type 1 Diabetes Genetics Consortium (T1DGC) reference panel ($n=5,225$), as well as the mixture of those two panels ($n=1,060$). HLARIMNT achieved higher accuracy than DEEP*HLA by several indices, especially for infrequent alleles. We also varied the size of data used for training, and HLARIMNT imputed more accurately among any size of training data. These results suggest that Transformer-based model may impute efficiently not only HLA types but also any other gene types from sequential SNPs data.\\
Relevant codes are available at  \url{https://github.com/seitalab/HLARIMNT}.

\end{abstract}

\section{Introduction}
\label{sec:intro}
The major histocompatibility complex (MHC) region is located on the short arm of chromosome 6, and is strongly associated with complex human traits \citep{hla_risk1}. It has been also found that Human leukocyte antigen (HLA) genes, which are abundant within the MHC region, explain much of the risk there \citep{hla_risk1}. Indeed, some HLA alleles are known to be at risk for the development of serious diseases, such as adverse reactions to drugs \citep{apli1,apli2}. Therefore, HLA genotyping is important in medicine.\\
\indent However, it is quite difficult to directly type the HLA alleles due to the complexity of the MHC region. Sanger sequencing and next-generation sequencing (NGS) are commonly used for direct typing of alleles, but these methods are time-consuming, expensive, and not suitable for mass production of analysis results. Furthermore, the limitation in terms of HLA gene coverage and allele resolution make it more difficult to type alleles \citep{diff_in_seq2,diff_in_seq1}. \\
\indent For these reasons, HLA alleles are usually computationally imputed by statistical models, based on observed single nucleotide polymorphism (SNP) data from ethnicity-specific reference panels \citep{diff_in_seq2,imputation4,imputation5,imputation6}.
For example, HLA*IMP is one of the imputation methods based on the Li \& Stephens haplotype model \citep{li_stephen_classical_model} using SNP data from European populations \citep{hla_imp1,hla_imp2}. HLA*IMP:02 is the subsequently developed method, which uses SNP data from multiple populations \citep{hlaimp02}. SNP2HLA, which uses the imputation software package Beagle to impute classical HLA alleles, is also one of the tools with high accuracy \citep{snp2hla}. HLA Genotype Imputation with Attribute Bagging (HIBAG), which is a method using multiple expectation–maximization-based classifiers, estimates the likelihood of HLA alleles \citep{hibag}. CookHLA is based on the standard hidden Markov model that can incorporate the genetic distance as input, using Beagle v4 and v5 \citep{cook_hla}.\\
\indent Nevertheless, there was still room for improvement in these imputation methods, in terms of imputation accuracy, especially for infrequent alleles. Since reference panels were used directly  other than HIBAG, there were restrictions on the data that can be accessed from the standpoint of personal information protection, which further reduced the accuracy of the imputation.
\\
\indent Turning to the area of machine learning, deep learning has made great strides in various domains. In addition to classical models like CNN and RNN, Transformer boasts high accuracy \citep{transformer}. Taking advantage of the attention mechanism and positional encoding, Transformer is good at handling sequential data, e.g. natural language \citep{transformer}. Now Transformer has been applied not only to natural language processing \citep{nlp1,nlp2}, but also to image recognition \citep{vit}, predictions of protein structures \citep{alpha_fold}, music generation \citep{music_transformer}, and image generation \citep{img_rcg1,img_rcg2}, by processing data as sequential. Deep learning models like these have also been applied to the field of medicine \citep{dlinmed,ml4h_lastyear}, including genetics \citep{binding,rnn_imputation}.
DEEP*HLA, which is a CNN-based model, was developed for HLA imputation \citep{deep_hla}. DEEP*HLA was a great advance in the field of HLA imputation in that it allowed more accurate imputation than existing methods, without the need to use a direct reference panel. Nevertheless, there was still room for improvement in DEEP*HLA, especially in accuracy for infrequent alleles, and  a large size of reference panel required for efficient imputation.\\ 
\indent In this study, we propose a Transformer-based model to impute eight HLA classical alleles, named "HLA Reliable IMputatioN by Transformer (HLARIMNT)", which allows the Transformer to take advantage of the sequential nature of SNPs. HLARIMNT performed imputations with generally higher accuracy than DEEP*HLA.
\section{Methods}
\label{sec:methods}
\begin{figure*}[t]
 \begin{center}
  \includegraphics[width=0.63\linewidth]{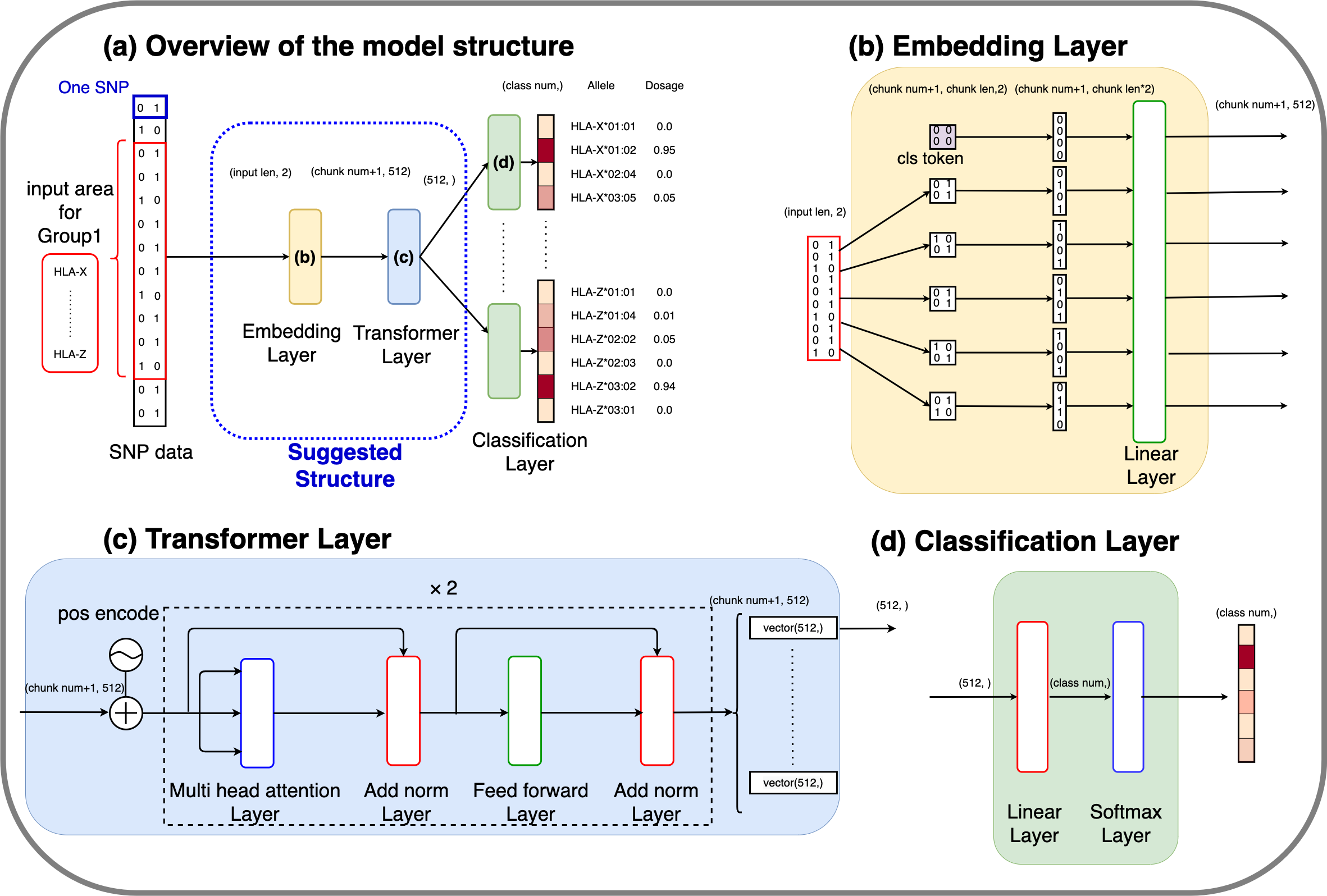}
  \caption{Architectures of HLARIMNT.\quad (a) overview of the model structure. (b) Embedding Layer. (c) Transformer Layer. (d) Classification Layer. HLARIMNT takes the input of each haplotype SNP genotype from pre-phased data represented as binary vectors, and outputs the genotype dosages of alleles for each HLA gene. input len: the number of SNPs to be input in Embedding Layer; chunk num: the number of chunks into which the SNPs are divided in Embedding Layer; chunk len: the number of SNPs in a single chunk; class num: the dimension of the output of Classification Layer.}
  \label{fig:architectures}
 \end{center}
\end{figure*}
\paragraph{Architectures.}
\label{subsec:architecture}
We modified the CNN part of DEEP*HLA into Transformer-based model (the blue dotted line in Figure \ref{fig:architectures}, i.e., Embedding Layer and Transformer Layer). Specific parameters and their values are listed in Appendix \ref{apd:hyper}.\\
\indent HLA genes are divided into groups according to LD structure and physical distance as in DEEP*HLA; (1){HLA-A}, (2){HLA-C, HLA-B}, (3){HLA-DRB1, HLA-DQA1, HLA-DQB1} and (4){HLA-DPA1, HLA-DPB1}, and genes in each group are imputed simultaneously. Therefore, the hyperparameters of the models are the same except for the number of outputs, but the weights trained are different for each group. For each group, the model takes the input of each haplotype SNP genotype from pre-phased data, which are expressed by two-dimensional vectors. The SNPs are expressed by 01 or 10 based on whether each base is consistent with a reference or alternative one. The range of SNPs used for training in each group is the same as DEEP*HLA; 500kbps each.\\
\indent Embedding Layer (Figure \ref{fig:architectures} (b)) consolidates the neighboring SNPs together to divide them into 50 chunks, and adds a classification (cls) token at the head of SNPs to learn the features. The cls token has the same shape as a single chunk, and its elements are all zeros. Then a common linear layer projects SNPs to the size of (51, 512). \\
\indent Transformer Layer (Figure \ref{fig:architectures} (c)) applies positional encoding and the encoder portion of Transformer to the data, after which the feature vector of cls token is extracted. \\
\indent Classification Layer (Figure \ref{fig:architectures} (d)), which is prepared for each HLA gene, is a combination of a linear layer and a softmax layer. Output dimension of the layer is the same as the number of alleles the HLA gene has. Output values are imputation dosages for the alleles, each of which takes a value from 0.0 to 1.0 that should be 1.0 when summed up.
\paragraph{Datasets.}
\label{subsec:datasets}
We used Pan-Asian reference panel \citep{pan_asian_ref_panel1,pan_asian_ref_panel2} and Type 1 Diabetes Genetics Consortium (T1DGC) reference panel \citep{t1dgc}. Both panels were genotyped using Illumina Immunochip and contain 4-digit resolution typing data of eight classical HLA genes based on SSO method; HLA-A, HLA-B, HLA-C, HLA-DRB1, HLA-DQA1, HLA-DQB1, HLA-DPA1, and HLA-DPB1. Both were distributed with Beagle format in a phased condition. Pan-Asian reference panel contains 530 unrelated individuals, i.e. 1060 haplotypes, of Asian ancestry. T1DGC reference panel contains 5225 unrelated individuals, i.e.10450 haplotypes, of European ancestry.
\paragraph{Training.}
\label{subsec:training}
In the training, as with DEEP*HLA, we adopted hierarchical fine-tuning, in which the parameters for classifying 2-digit alleles were transferred to the model for 4-digit alleles. This allowed the model to take advantage of the hierarchical nature of HLA alleles. 
We only used SNP data for training and evaluation, and removed other information from the reference panel. We used Cross Entropy Loss as a loss function and Adam to optimize the loss. The Cross Entropy Loss is expressed by the following equation, where $t_i$ indicates correct output that should be 0 or 1, $y_i$ indicates the output of softmax layer, and $n$ indicates the number of classes. 
\begin{align}
    Loss = -\sum_{i=1}^{n} t_i\log{y_i}\notag
\end{align}
We used 10 percent of the training data as validation data for early stopping and model updates during the training.
The training flow is the same as in DEEP*HLA (Appendix \ref{apd:train_flow}). 
\paragraph{Evaluation.}
\label{subsec:evaluation}
 In the experiments, we used 4 indices for the evaluations; $r^2$, PPV, sensitivity, and probability, values of which were calculated for each allele.\\ 
\indent $r^2(A)$ represents Pearson’s product moment correlation coefficient between imputed and typed dosages and is expressed as follows; 
\begin{figure}[h]
 \begin{center}
  \includegraphics[width=1.0\linewidth]{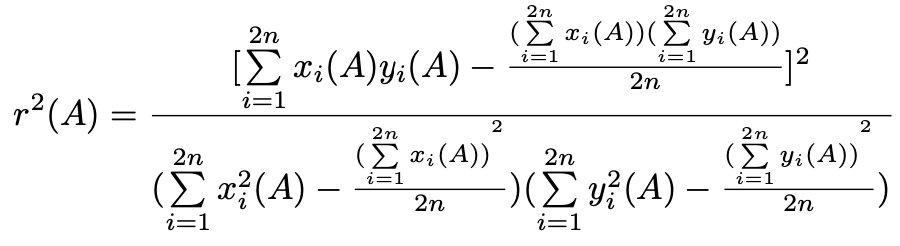}
 \end{center}

\end{figure}
\\where $n$ is the number of individuals (i.e.$2n$ is the number of haplotypes), $A$ is the type of allele (eg. HLA-A*01:23), $x_i(A)$ is the imputed dosage of allele$A$ for haplotype $i$ , which is obtained from the output of the softmax layer, and $y_i(A)$ is the typed dosage of allele$A$ for haplotype $i$, taking values 1 if haplotype $i$ has allele$A$ and otherwise 0.\\
\indent Sensitivity($A$) represents the percentage of haplotypes that were correctly imputed to have allele$A$ in all the haplotypes that have allele$A$, and PPV($A$) represents the percentage of haplotypes with allele$A$ in all the haplotypes predicted to have allele$A$ \citep{deep_hla}. \\
\indent Probability($A$) represents the imputed dosage of allele$A$ for each haplotype with allele$A$, and is expressed as follows;
\begin{align}
    probability(A)=\sum_{i=1}^{m} x_i(A)/m\notag
\end{align}
where $m$ represents the number of haplotypes with allele$A$.

\section{Results}
\subsection{Experiment 1}
\label{subsec:exp1}
We first performed a five-part cross-validation for Pan-Asian reference panel, T1DGC reference panel, and the mixture of the two panels (\ref{subsec:datasets}/Datasets). 
When creating the mixed panel, we randomly selected the same number of individuals included in Pan-Asian reference panel from T1DGC counterpart.
Thus, the number of individuals included in the mixed panel became 1060. We divided the data into 5 parts for the cross-validation and used 10 percent of the non-test data as validation data during the training (\ref{subsec:training}/Training).\\
Using HLARIMNT and DEEP*HLA, we performed imputations of 4-digit alleles for eight HLA genes (\ref{subsec:datasets}/Datasets). Then, we calculated the accuracy for each allele as four indices; $r^2$, PPV, sensitivity, and probability (\ref{subsec:evaluation}/Evaluation). Finally we compared the average of each indice between the two methods (\ref{subsec:architecture}/Architectures).

\paragraph{HLARIMNT performed better in general.}

\begin{table}[h]\centering
\caption{\bfseries Average imputation accuracy for all 4-digit alleles}
\label{table:table1}
\small
\scalebox{0.6}{
\begin{tabular}{c|c|c|c}
\bfseries Dataset&\bfseries Indices&\bfseries DEEP*HLA&\bfseries HLARIMNT\\ \hline
\bfseries Pan-Asian&$\mathbf r^2$ &0.795&\bfseries0.850\\ 
&\bfseries PPV&0.844&\bfseries0.865\\ 
&\bfseries sensitivity&0.773&\bfseries0.820\\ 
&\bfseries probability&0.771&\bfseries0.784\\ \hline
\bfseries T1DGC&$\mathbf r^2$ &0.823& \bfseries0.839 \\ 
&\bfseries PPV&\bfseries0.897&0.881\\ 
&\bfseries sensitivity&0.786&\bfseries0.804\\ 
&\bfseries probability&0.786&\bfseries0.794\\ \hline
\bfseries Mixed&$\mathbf r^2$ & 0.795&\bfseries 0.830\\ 
&\bfseries PPV&0.862&\bfseries0.870\\ 
&\bfseries sensitivity&0.761&\bfseries0.797\\ 
&\bfseries probability&0.761&\bfseries0.778\\ \hline
\end{tabular}
}
\end{table}

Table \ref{table:table1} shows the average imputation accuracy of all 4-digit alleles. Confidence intervals for the each test data are provided in Appendix \ref{apd:interval_1} and the weighted average values by allele frequency are provided in Appendix \ref{apd:weighted}. \\
\indent HLARIMNT achieved higher accuracy in the 4 indices on all the reference panels, although PPV in T1DGC reference panel was a little higher for DEEP*HLA.\\
\indent Average imputation accuracy for 4-digit alleles in each HLA gene is showed in Figure \ref{fig:fig3} of Appendix \ref{apd:exp1}. HLARIMNT generally showed advantages for almost all of the genes in all of the three reference panels, with some exceptions depending on the combination of genes, reference panels, and indices.
\paragraph{HLARIMNT outperformed for infrequent alleles.}
\begin{table}[h]\centering
\caption{\bfseries Average imputation accuracy for infrequent 4-digit alleles}
\label{table:table2}
\small
\scalebox{0.6}{
\begin{tabular}{c|c|c|c}
\bfseries Dataset&\bfseries Indices&\bfseries DEEP*HLA&\bfseries HLARIMNT\\ \hline
\bfseries Pan-Asian&$\mathbf r^2$ &0.583 &\bfseries0.685\\ 
&\bfseries PPV&\bfseries0.740&0.725\\ 
&\bfseries sensitivity&0.447&\bfseries0.553\\ 
&\bfseries probability&0.446&\bfseries0.502\\ \hline
\bfseries T1DGC&$\mathbf r^2$ &0.715 &\bfseries0.740\\ 
&\bfseries PPV&\bfseries0.834&0.805\\ 
&\bfseries sensitivity&0.649&\bfseries0.678\\ 
&\bfseries probability&0.649&\bfseries0.665\\ \hline
\bfseries Mixed&$\mathbf r^2$ &0.628&\bfseries0.681\\ 
&\bfseries PPV&0.762&\bfseries0.768\\ 
&\bfseries sensitivity&0.531&\bfseries0.600\\ 
&\bfseries probability&0.530&\bfseries0.577\\ \hline

\end{tabular}
}
\end{table}

Table \ref{table:table2} shows the average imputation accuracy for 4-digit alleles that have frequencies less than 0.01. Confidence intervals for each test data are provided in Appendix \ref{apd:interval_2} and the weighted average values by allele frequency are listed in Appendix \ref{apd:weighted}. As for all alleles, HLARIMNT outperformed the accuracy of DEEP*HLA, except for PPV of Pan-Asian and T1DGC reference penel. Even more noteworthy is the magnitude of the difference in accuracy between the two methods. The superiority of HLARIMNT for infrequent alleles was greater than that of for all alleles.
\\
\indent Average imputation accuracy of 4-digit alleles for each allele frequency is showed in Figure \ref{fig:fig4} of Appendix \ref{apd:exp1}. The superiority of HLARIMNT was more noticeable for alleles with lower frequencies.

\subsection{Experiment 2}
\label{subsec:exp2}
We performed a five-part cross-validation for T1DGC reference panel, the training data of which consisted of various numbers of individuals; 530, 1300, 2600, and 4180. We randomly selected data of these numbers from T1DGC reference panel for training. The test data used for evaluation was the same regardless of the training data size. The data splitting for this experiment is described in detail in Appendix \ref{apd:third}. We used the same four indices as in Experiment 1 for the evaluation.

\paragraph{HLARIMNT generally performed better in any training data numbers.}
\label{subsubsec:general}
\begin{figure}[h]
 \begin{center}
  \includegraphics[width=0.77\linewidth]{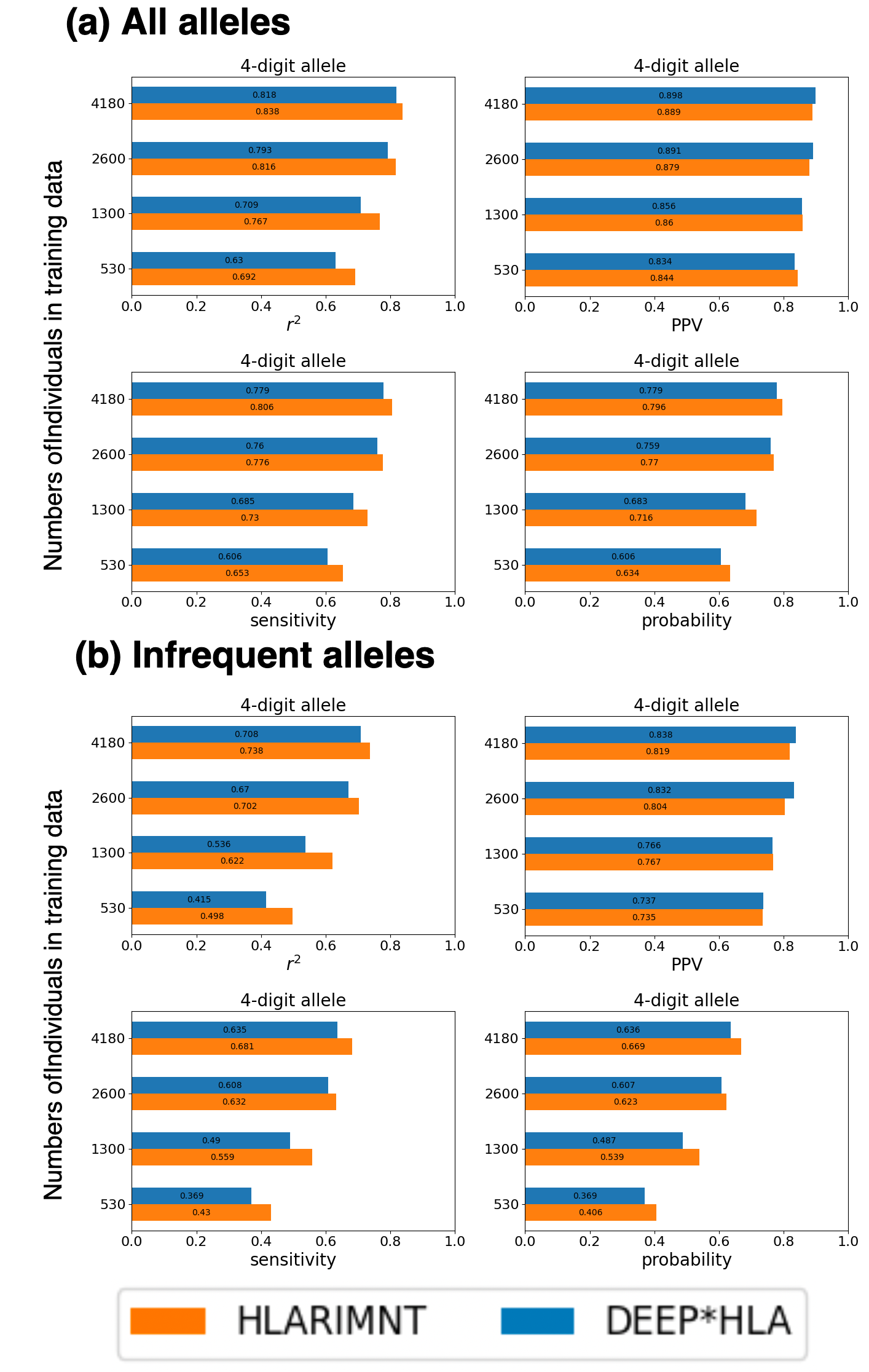}
  \caption{(a) shows the average accuracy for 4-digit alleles using various numbers of training data and (b) shows that of alleles which have frequencies less than 0.01. The accuracy of HLARIMNT was generally higher than that of DEEP*HLA on any training data numbers, especially on small numbers.}
  \label{fig:fig6}
 \end{center}
\end{figure}
Figure \ref{fig:fig6} (a) shows the average accuracy for 4-digit alleles using various numbers of training data. The numbers in the bar graphs are the accuracy values that the graphs represent. The accuracy of HLARIMNT was higher than that of DEEP*HLA regardless of training data numbers, with the exception of PPV using relatively large sizes of training data. In addition, in all of the indices, the superiority of HLARIMNT to DEEP*HLA was more noticeable when the number of training data was smaller, although not completely in order.
Even in PPV, HLARIMNT was superior to DEEP*HLA when the number of training data was relatively small.

\paragraph{HLARIMNT performed better for infrequent alleles in any training data numbers.}
Figure \ref{fig:fig6} (b) shows the average accuracy for 4-digit alleles with frequencies less than 0.01. As with the previous results (\ref{subsubsec:general}), HLARIMNT outperformed DEEP*HLA in all of the training data numbers, except for PPV. Also, there was the tendency that the advantage of HLARIMNT was more pronounced when the number of data for training was smaller.

\section{Conclusion}
As shown in Experiment 1 (\ref{subsec:exp1}), HLARIMNT achieved higher accuracy than DEEP*HLA in all of the three reference panels, especially for infrequent alleles. In addition, as shown in Experiment 2 (\ref{subsec:exp2}), HLARIMNT outperformed regardless of the size of data used for training.\\
\indent The results of these two experiments suggest that HLARIMNT can perform HLA imputation efficiently, regardless of race and the size of available data. Further discussion is provided in Appendix \ref{apd:discussion}.
\section{Data availability and ethical review}
We used two HLA reference panels in this study. Both panels were distributed as a phased condition, and publicly available. Pan-Asian reference panel is downloadable with SNP2HLA software (\url{http://software.broadinstitute.org/mpg/snp2hla/}). T1DGC panel is available from the NIDDK Central Repository after the registration process (\url{https://repository.niddk.nih.gov/studies/t1dgc/}). Since all data used in this study have already been published and are publicly available, the ethics committees of our institute have waived an ethical review.

\clearpage
\bibliography{pmlr-sample}

\begin{thebibliography}{37}
\providecommand{\natexlab}[1]{#1}
\providecommand{\url}[1]{\texttt{#1}}
\expandafter\ifx\csname urlstyle\endcsname\relax
  \providecommand{\doi}[1]{doi: #1}\else
  \providecommand{\doi}{doi: \begingroup \urlstyle{rm}\Url}\fi

\bibitem[Adiwardana et~al.(2020)Adiwardana, Luong, So, Hall, Fiedel, Thoppilan,
  Yang, Kulshreshtha, Nemade, Lu, and Le]{nlp2}
Daniel Adiwardana, Minh-Thang Luong, David~R. So, Jamie Hall, Noah Fiedel,
  Romal Thoppilan, Zi~Yang, Apoorv Kulshreshtha, Gaurav Nemade, Yifeng Lu, and
  Quoc~V. Le.
\newblock Towards a human-like open-domain chatbot.
\newblock 1 2020.
\newblock URL \url{http://arxiv.org/abs/2001.09977}.

\bibitem[Brown et~al.(2020)Brown, Mann, Ryder, Subbiah, Kaplan, Dhariwal,
  Neelakantan, Shyam, Sastry, Askell, Agarwal, Herbert-Voss, Krueger, Henighan,
  Child, Ramesh, Ziegler, Wu, Winter, Hesse, Chen, Sigler, Litwin, Gray, Chess,
  Clark, Berner, McCandlish, Radford, Sutskever, and Amodei]{nlp1}
Tom~B. Brown, Benjamin Mann, Nick Ryder, Melanie Subbiah, Jared Kaplan,
  Prafulla Dhariwal, Arvind Neelakantan, Pranav Shyam, Girish Sastry, Amanda
  Askell, Sandhini Agarwal, Ariel Herbert-Voss, Gretchen Krueger, Tom Henighan,
  Rewon Child, Aditya Ramesh, Daniel~M. Ziegler, Jeffrey Wu, Clemens Winter,
  Christopher Hesse, Mark Chen, Eric Sigler, Mateusz Litwin, Scott Gray,
  Benjamin Chess, Jack Clark, Christopher Berner, Sam McCandlish, Alec Radford,
  Ilya Sutskever, and Dario Amodei.
\newblock Language models are few-shot learners.
\newblock 5 2020.
\newblock URL \url{http://arxiv.org/abs/2005.14165}.

\bibitem[Cho et~al.(2014)Cho, van Merrienboer, Gulcehre, Bahdanau, Bougares,
  Schwenk, and Bengio]{gru}
Kyunghyun Cho, Bart van Merrienboer, Caglar Gulcehre, Dzmitry Bahdanau, Fethi
  Bougares, Holger Schwenk, and Yoshua Bengio.
\newblock Learning phrase representations using rnn encoder-decoder for
  statistical machine translation.
\newblock 6 2014.
\newblock URL \url{http://arxiv.org/abs/1406.1078}.

\bibitem[Chu et~al.(2022)Chu, Zhang, Wang, Zhang, Wang, Wang, Salahub, Xu,
  Wang, Jiang, Xiong, and Wei]{binding}
Yanyi Chu, Yan Zhang, Qiankun Wang, Lingfeng Zhang, Xuhong Wang, Yanjing Wang,
  Dennis~Russell Salahub, Qin Xu, Jianmin Wang, Xue Jiang, Yi~Xiong, and
  Dong~Qing Wei.
\newblock A transformer-based model to predict peptide–hla class i binding
  and optimize mutated peptides for vaccine design.
\newblock \emph{Nature Machine Intelligence}, 4:\penalty0 300--311, 3 2022.
\newblock ISSN 25225839.
\newblock \doi{10.1038/s42256-022-00459-7}.

\bibitem[Cook et~al.(2021)Cook, Choi, Lim, Luo, Kim, Jia, Raychaudhuri, and
  Han]{cook_hla}
Seungho Cook, Wanson Choi, Hyunjoon Lim, Yang Luo, Kunhee Kim, Xiaoming Jia,
  Soumya Raychaudhuri, and Buhm Han.
\newblock Accurate imputation of human leukocyte antigens with cookhla.
\newblock \emph{Nature Communications}, 12, 12 2021.
\newblock ISSN 20411723.
\newblock \doi{10.1038/s41467-021-21541-5}.

\bibitem[Dendrou et~al.(2018)Dendrou, Petersen, Rossjohn, and
  Fugger]{hla_risk1}
Calliope~A. Dendrou, Jan Petersen, Jamie Rossjohn, and Lars Fugger.
\newblock Hla variation and disease, 5 2018.
\newblock ISSN 14741741.

\bibitem[Dilthey et~al.(2013)Dilthey, Leslie, Moutsianas, Shen, Cox, Nelson,
  and McVean]{hlaimp02}
Alexander Dilthey, Stephen Leslie, Loukas Moutsianas, Judong Shen, Charles Cox,
  Matthew~R. Nelson, and Gil McVean.
\newblock Multi-population classical hla type imputation.
\newblock \emph{PLoS Computational Biology}, 9, 2 2013.
\newblock ISSN 1553734X.
\newblock \doi{10.1371/journal.pcbi.1002877}.

\bibitem[Dilthey et~al.(2011)Dilthey, Moutsianas, Leslie, and McVean]{hla_imp1}
Alexander~T. Dilthey, Loukas Moutsianas, Stephen Leslie, and Gil McVean.
\newblock Hla*imp-an integrated framework for imputing classical hla alleles
  from snp genotypes.
\newblock \emph{Bioinformatics}, 27:\penalty0 968--972, 4 2011.
\newblock ISSN 13674803.
\newblock \doi{10.1093/bioinformatics/btr061}.

\bibitem[Dosovitskiy et~al.(2020)Dosovitskiy, Beyer, Kolesnikov, Weissenborn,
  Zhai, Unterthiner, Dehghani, Minderer, Heigold, Gelly, Uszkoreit, and
  Houlsby]{vit}
Alexey Dosovitskiy, Lucas Beyer, Alexander Kolesnikov, Dirk Weissenborn,
  Xiaohua Zhai, Thomas Unterthiner, Mostafa Dehghani, Matthias Minderer, Georg
  Heigold, Sylvain Gelly, Jakob Uszkoreit, and Neil Houlsby.
\newblock An image is worth 16x16 words: Transformers for image recognition at
  scale.
\newblock 10 2020.
\newblock URL \url{http://arxiv.org/abs/2010.11929}.

\bibitem[Egger et~al.(2022)Egger, Gsaxner, Pepe, Pomykala, Jonske, Kurz, Li,
  and Kleesiek]{dlinmed}
Jan Egger, Christina Gsaxner, Antonio Pepe, Kelsey~L. Pomykala, Frederic
  Jonske, Manuel Kurz, Jianning Li, and Jens Kleesiek.
\newblock Medical deep learning—a systematic meta-review.
\newblock \emph{Computer Methods and Programs in Biomedicine}, 221:\penalty0
  106874, 6 2022.
\newblock ISSN 01692607.
\newblock \doi{10.1016/j.cmpb.2022.106874}.

\bibitem[Erlich(2012)]{diff_in_seq1}
H.~Erlich.
\newblock Hla dna typing: Past, present, and future, 7 2012.
\newblock ISSN 00012815.

\bibitem[Gourraud et~al.(2014)Gourraud, Khankhanian, Cereb, Yang, Feolo,
  Maiers, Rioux, Hauser, and Oksenberg]{distri_by_eth}
Pierre~Antoine Gourraud, Pouya Khankhanian, Nezih Cereb, Soo~Young Yang,
  Michael Feolo, Martin Maiers, John~D. Rioux, Stephen Hauser, and Jorge
  Oksenberg.
\newblock Hla diversity in the 1000 genomes dataset.
\newblock \emph{PLoS ONE}, 9, 7 2014.
\newblock ISSN 19326203.
\newblock \doi{10.1371/journal.pone.0097282}.

\bibitem[Halevy et~al.(2008)Halevy, Ghislain, Mockenhaupt, Fagot, {Bouwes
  Bavinck}, Sidoroff, Naldi, Dunant, Viboud, and Roujeau]{apli1}
Sima Halevy, Pierre-Dominique Ghislain, Maja Mockenhaupt, Jean-Paul Fagot,
  Jan~Nico {Bouwes Bavinck}, Alexis Sidoroff, Luigi Naldi, Ariane Dunant,
  Cecile Viboud, and Jean-Claude Roujeau.
\newblock Allopurinol is the most common cause of stevens-johnson syndrome and
  toxic epidermal necrolysis in europe and israel.
\newblock \emph{Journal of the American Academy of Dermatology}, 58\penalty0
  (1):\penalty0 25--32, 2008.
\newblock ISSN 0190-9622.
\newblock \doi{https://doi.org/10.1016/j.jaad.2007.08.036}.
\newblock URL
  \url{https://www.sciencedirect.com/science/article/pii/S0190962207013217}.

\bibitem[Han et~al.(2014)Han, Diogo, Eyre, Kallberg, Zhernakova, Bowes,
  Padyukov, Okada, González-Gay, Rantapää-Dahlqvist, Martin, Huizinga,
  Plenge, Worthington, Gregersen, Klareskog, Bakker, and Raychaudhuri]{t1dgc}
Buhm Han, Dorothée Diogo, Steve Eyre, Henrik Kallberg, Alexandra Zhernakova,
  John Bowes, Leonid Padyukov, Yukinori Okada, Miguel~A. González-Gay,
  Solbritt Rantapää-Dahlqvist, Javier Martin, Tom~W.J. Huizinga, Robert~M.
  Plenge, Jane Worthington, Peter~K. Gregersen, Lars Klareskog, Paul I.W.~De
  Bakker, and Soumya Raychaudhuri.
\newblock Fine mapping seronegative and seropositive rheumatoid arthritis to
  shared and distinct hla alleles by adjusting for the effects of
  heterogeneity.
\newblock \emph{American Journal of Human Genetics}, 94:\penalty0 522--532, 4
  2014.
\newblock ISSN 15376605.
\newblock \doi{10.1016/j.ajhg.2014.02.013}.

\bibitem[Hirata et~al.(2019)Hirata, Hosomichi, Sakaue, Kanai, Nakaoka,
  Ishigaki, Suzuki, Akiyama, Kishikawa, Ogawa, Masuda, Yamamoto, Hirata,
  Matsuda, Momozawa, Inoue, Kubo, Kamatani, and Okada]{diff_in_seq2}
Jun Hirata, Kazuyoshi Hosomichi, Saori Sakaue, Masahiro Kanai, Hirofumi
  Nakaoka, Kazuyoshi Ishigaki, Ken Suzuki, Masato Akiyama, Toshihiro Kishikawa,
  Kotaro Ogawa, Tatsuo Masuda, Kenichi Yamamoto, Makoto Hirata, Koichi Matsuda,
  Yukihide Momozawa, Ituro Inoue, Michiaki Kubo, Yoichiro Kamatani, and
  Yukinori Okada.
\newblock Genetic and phenotypic landscape of the major histocompatibilty
  complex region in the japanese population.
\newblock \emph{Nature Genetics}, 51:\penalty0 470--480, 3 2019.
\newblock ISSN 15461718.
\newblock \doi{10.1038/s41588-018-0336-0}.

\bibitem[Hu et~al.(2015)Hu, Deutsch, Lenz, Onengut-Gumuscu, Han, Chen, Howson,
  Todd, Bakker, Rich, and Raychaudhuri]{dqbeta_europe2}
Xinli Hu, Aaron~J. Deutsch, Tobias~L. Lenz, Suna Onengut-Gumuscu, Buhm Han,
  Wei~Min Chen, Joanna~M.M. Howson, John~A. Todd, Paul I.W.~De Bakker,
  Stephen~S. Rich, and Soumya Raychaudhuri.
\newblock Additive and interaction effects at three amino acid positions in
  hla-dq and hla-dr molecules drive type 1 diabetes risk.
\newblock \emph{Nature Genetics}, 47:\penalty0 898--905, 8 2015.
\newblock ISSN 15461718.
\newblock \doi{10.1038/ng.3353}.

\bibitem[Huang et~al.(2018)Huang, Vaswani, Uszkoreit, Shazeer, Simon,
  Hawthorne, Dai, Hoffman, Dinculescu, and Eck]{music_transformer}
Cheng-Zhi~Anna Huang, Ashish Vaswani, Jakob Uszkoreit, Noam Shazeer, Ian Simon,
  Curtis Hawthorne, Andrew~M. Dai, Matthew~D. Hoffman, Monica Dinculescu, and
  Douglas Eck.
\newblock Music transformer.
\newblock 9 2018.
\newblock URL \url{http://arxiv.org/abs/1809.04281}.

\bibitem[Jia et~al.(2013)Jia, Han, Onengut-Gumuscu, Chen, Concannon, Rich,
  Raychaudhuri, and de~Bakker]{snp2hla}
Xiaoming Jia, Buhm Han, Suna Onengut-Gumuscu, Wei~Min Chen, Patrick~J.
  Concannon, Stephen~S. Rich, Soumya Raychaudhuri, and Paul~I.W. de~Bakker.
\newblock Imputing amino acid polymorphisms in human leukocyte antigens.
\newblock \emph{PLoS ONE}, 8, 6 2013.
\newblock ISSN 19326203.
\newblock \doi{10.1371/journal.pone.0064683}.

\bibitem[Jumper et~al.(2021)Jumper, Evans, Pritzel, Green, Figurnov,
  Ronneberger, Tunyasuvunakool, Bates, Žídek, Potapenko, Bridgland, Meyer,
  Kohl, Ballard, Cowie, Romera-Paredes, Nikolov, Jain, Adler, Back, Petersen,
  Reiman, Clancy, Zielinski, Steinegger, Pacholska, Berghammer, Bodenstein,
  Silver, Vinyals, Senior, Kavukcuoglu, Kohli, and Hassabis]{alpha_fold}
John Jumper, Richard Evans, Alexander Pritzel, Tim Green, Michael Figurnov,
  Olaf Ronneberger, Kathryn Tunyasuvunakool, Russ Bates, Augustin Žídek, Anna
  Potapenko, Alex Bridgland, Clemens Meyer, Simon~A.A. Kohl, Andrew~J. Ballard,
  Andrew Cowie, Bernardino Romera-Paredes, Stanislav Nikolov, Rishub Jain,
  Jonas Adler, Trevor Back, Stig Petersen, David Reiman, Ellen Clancy, Michal
  Zielinski, Martin Steinegger, Michalina Pacholska, Tamas Berghammer,
  Sebastian Bodenstein, David Silver, Oriol Vinyals, Andrew~W. Senior, Koray
  Kavukcuoglu, Pushmeet Kohli, and Demis Hassabis.
\newblock Highly accurate protein structure prediction with alphafold.
\newblock \emph{Nature}, 596:\penalty0 583--589, 8 2021.
\newblock ISSN 14764687.
\newblock \doi{10.1038/s41586-021-03819-2}.

\bibitem[Kawabata et~al.(2009)Kawabata, Ikegami, Awata, Imagawa, Maruyama,
  Kawasaki, Tanaka, Shimada, Osawa, Kobayashi, Hanafusa, Tokunaga, and
  Makino]{dqbeta_jap}
Y.~Kawabata, H.~Ikegami, T.~Awata, A.~Imagawa, T.~Maruyama, E.~Kawasaki,
  S.~Tanaka, A.~Shimada, H.~Osawa, T.~Kobayashi, T.~Hanafusa, K.~Tokunaga, and
  H.~Makino.
\newblock Differential association of hla with three subtypes of type 1
  diabetes: Fulminant, slowly progressive and acute-onset.
\newblock \emph{Diabetologia}, 52:\penalty0 2513--2521, 12 2009.
\newblock ISSN 0012186X.
\newblock \doi{10.1007/s00125-009-1539-9}.

\bibitem[Ko et~al.(2015)Ko, Tsai, Chen, Chen, Yu, Chu, Huang, Wang, Weng, Yu,
  Hsieh, Tsai, Lai, Tsai, Yin, Ou, Cheng, Yen, Liou, Lin, Chen, Hsiao, Weng,
  Chen, Chen, Liu, Yen, Lee, Kuo, Wu, Hung, Luo, Yang, Chuang, Chou, Liao,
  Wang, Huang, Chang, Lee, Chen, Wong, Chen, Wu, Chen, and Shen]{apli2}
Tai~Ming Ko, Chang~Youh Tsai, Shih~Yang Chen, Kuo~Shu Chen, Kuang~Hui Yu,
  Chih~Sheng Chu, Chung~Ming Huang, Chrong~Reen Wang, Chia~Tse Weng, Chia~Li
  Yu, Song~Chou Hsieh, Jer~Chia Tsai, Wen~Ter Lai, Wen~Chan Tsai, Guang~Dar
  Yin, Tsan~Teng Ou, Kai~Hung Cheng, Jeng~Hsien Yen, Teh~Ling Liou, Tsung~Hsien
  Lin, Der~Yuan Chen, Pi~Jung Hsiao, Meng~Yu Weng, Yi~Ming Chen, Chen~Hung
  Chen, Ming~Fei Liu, Hsueh~Wei Yen, Jia~Jung Lee, Mei~Chuan Kuo, Chen~Ching
  Wu, Shih~Yuan Hung, Shue~Fen Luo, Ya~Hui Yang, Hui~Ping Chuang, Yi~Chun Chou,
  Hung~Ting Liao, Chia~Wen Wang, Chun~Lin Huang, Chia~Shuo Chang, Ming
  Ta~Michael Lee, Pei Chen, Chih~Shung Wong, Chien~Hsiun Chen, Jer~Yuarn Wu,
  Yuan~Tsong Chen, and Chen~Yang Shen.
\newblock Use of hla-b 58:01$*$genotyping to prevent allopurinol induced severe
  cutaneous adverse reactions in taiwan: National prospective cohort study.
\newblock \emph{BMJ (Online)}, 351, 9 2015.
\newblock ISSN 17561833.
\newblock \doi{10.1136/bmj.h4848}.

\bibitem[Kojima et~al.(2020)Kojima, Tadaka, Katsuoka, Tamiya, Yamamoto, and
  Kinoshita]{rnn_imputation}
Kaname Kojima, Shu Tadaka, Fumiki Katsuoka, Gen Tamiya, Masayuki Yamamoto, and
  Kengo Kinoshita.
\newblock A genotype imputation method for de-identified haplotype reference
  information by using recurrent neural network.
\newblock \emph{PLoS Computational Biology}, 16, 10 2020.
\newblock ISSN 15537358.
\newblock \doi{10.1371/journal.pcbi.1008207}.

\bibitem[Leslie et~al.(2008)Leslie, Donnelly, and
  McVean]{li_stephen_classical_model}
Stephen Leslie, Peter Donnelly, and Gil McVean.
\newblock A statistical method for predicting classical hla alleles from snp
  data.
\newblock \emph{American Journal of Human Genetics}, 82:\penalty0 48--56, 1
  2008.
\newblock ISSN 00029297.
\newblock \doi{10.1016/j.ajhg.2007.09.001}.

\bibitem[Li and Stephens()]{hla_imp2}
Na~Li and Matthew Stephens.
\newblock Modeling linkage disequilibrium and identifying recombination
  hotspots using single-nucleotide polymorphism data tering of recombination
  events in the ancestry of the.

\bibitem[Naito et~al.(2021)Naito, Suzuki, Hirata, Kamatani, Matsuda, Toda, and
  Okada]{deep_hla}
Tatsuhiko Naito, Ken Suzuki, Jun Hirata, Yoichiro Kamatani, Koichi Matsuda,
  Tatsushi Toda, and Yukinori Okada.
\newblock A deep learning method for hla imputation and trans-ethnic mhc
  fine-mapping of type 1 diabetes.
\newblock \emph{Nature Communications}, 12, 12 2021.
\newblock ISSN 20411723.
\newblock \doi{10.1038/s41467-021-21975-x}.

\bibitem[Okada et~al.(2014)Okada, Kim, Han, Pillai, Ong, Saw, Luo, Jiang, Yin,
  Bang, Lee, Brown, Bae, Xu, Teo, de~Bakker, and
  Raychaudhuri]{pan_asian_ref_panel2}
Yukinori Okada, Kwangwoo Kim, Buhm Han, Nisha~E. Pillai, Rick~T.H. Ong,
  Woei~Yuh Saw, Ma~Luo, Lei Jiang, Jian Yin, So~Young Bang, Hye~Soon Lee,
  Matthew~A. Brown, Sang~Cheol Bae, Huji Xu, Yik~Ying Teo, Paul~I.W. de~Bakker,
  and Soumya Raychaudhuri.
\newblock Risk for acpa-positive rheumatoid arthritis is driven by shared hla
  amino acid polymorphisms in asian and european populations.
\newblock \emph{Human molecular genetics}, 23:\penalty0 6916--6926, 12 2014.
\newblock ISSN 14602083.
\newblock \doi{10.1093/hmg/ddu387}.

\bibitem[Okada et~al.(2015)Okada, Momozawa, Ashikawa, Kanai, Matsuda, Kamatani,
  Takahashi, and Kubo]{imputation6}
Yukinori Okada, Yukihide Momozawa, Kyota Ashikawa, Masahiro Kanai, Koichi
  Matsuda, Yoichiro Kamatani, Atsushi Takahashi, and Michiaki Kubo.
\newblock Construction of a population-specific hla imputation reference panel
  and its application to graves' disease risk in japanese.
\newblock \emph{Nature Genetics}, 47:\penalty0 798--802, 6 2015.
\newblock ISSN 15461718.
\newblock \doi{10.1038/ng.3310}.

\bibitem[Pereyra et~al.(2010)Pereyra, Jia, McLaren, Telenti, de~Bakker, Walker,
  Ripke, Brumme, Pulit, Carrington, Kadie, Carlson, Heckerman, Graham, Plenge,
  Deeks, Gianniny, Crawford, Sullivan, Gonzalez, Davies, Camargo, Moore,
  Beattie, Gupta, Crenshaw, Burtt, Guiducci, Gupta, Gao, Qi, Yuki,
  Piechocka-Trocha, Cutrell, Rosenberg, Moss, Lemay, O’leary, Schaefer,
  Verma, Toth, Block, Baker, Rothchild, Lian, Proudfoot, Alvino, Vine, Addo,
  Allen, Altfeld, Henn, Gall, Streeck, Haas, Kuritzkes, Robbins, Shafer,
  Gulick, Shikuma, Haubrich, Riddler, Sax, Daar, Ribaudo, Agan, Agarwal, Ahern,
  Allen, Altidor, Altschuler, Ambardar, Anastos, Anderson, Anderson, Andrady,
  Antoniskis, Bangsberg, Barbaro, Barrie, Bartczak, Barton, Basden, Basgoz,
  Bazner, Bellos, Benson, Berger, Bernard, Bernard, Birch, Bodner, Bolan,
  Boudreaux, Bradley, Braun, Brndjar, Brown, Brown, Brown, Burack, Bush,
  Cafaro, Campbell, Campbell, Carlson, Carmichael, Casey, Cavacuiti, Celestin,
  Chambers, Chez, Chirch, Cimoch, Cohen, Cohn, Conway, Cooper, Cornelson, Cox,
  Cristofano, Cuchural, Czartoski, Dahman, Daly, Davis, Davis, Davod, Dejesus,
  Dietz, Dunham, Dunn, Ellerin, Eron, Fangman, Farel, Ferlazzo, Fidler,
  Fleenor-Ford, Frankel, Freedberg, French, Fuchs, Fuller, Gaberman, Gallant,
  Gandhi, Garcia, Garmon, Gathe, Gaultier, Gebre, Gilman, Gilson, Goepfert,
  Gottlieb, Goulston, Groger, Gurley, Haber, Hardwicke, Hardy, Harrigan,
  Hawkins, Heath, Hecht, Henry, Hladek, Hoffman, Horton, Hsu, Huhn, Hunt,
  Hupert, Illeman, Jaeger, Jellinger, John, Johnson, Johnson, Johnson, Johnson,
  Joly, Jordan, Kauffman, Khanlou, Killian, Kim, Kim, Kinder, Kirchner,
  Kogelman, Kojic, Korthuis, Kurisu, Kwon, Lamar, Lampiris, Lanzafame,
  Lederman, Lee, Lee, Lee, Lee, Lemoine, Levy, Llibre, Liguori, Little, Liu,
  Lopez, Loutfy, Loy, Mohammed, Man, Mansour, Marconi, Markowitz, Marques,
  Martin, Martin, Mayer, McElrath, McGhee, McGovern, McGowan, McIntyre, McLeod,
  Menezes, Mesa, Metroka, Meyer-Olson, Miller, Montgomery, Mounzer, Nagami,
  Nagin, Nahass, Nelson, Nielsen, Norene, O’connor, Ojikutu, Okulicz,
  Oladehin, Oldfield, Olender, Ostrowski, Owen, Pae, Parsonnet, Pavlatos,
  Perlmutter, Pierce, Pincus, Pisani, Price, Proia, Prokesch, Pujet, Ramgopal,
  Rathod, Rausch, Ravishankar, Rhame, Richards, Richman, Rodes, Rodriguez,
  Rose, Rosenberg, Rosenthal, Ross, Rubin, Rumbaugh, Saenz, Salvaggio, Sanchez,
  Sanjana, Santiago, Schmidt, Schuitemaker, Sestak, Shalit, Shay, Shirvani,
  Silebi, Sizemore, Skolnik, Sokol-Anderson, Sosman, Stabile, Stapleton,
  Starrett, Stein, Stellbrink, Sterman, Stone, Stone, Tambussi, Taplitz,
  Tedaldi, Theisen, Torres, Tosiello, Tremblay, Tribble, Trinh, Tsao, Ueda,
  Vaccaro, Valadas, Vanig, Vecino, Vega, Veikley, Wade, Walworth, Wanidworanun,
  Ward, Warner, Weber, Webster, Weis, Wheeler, White, Wilkins, Winston,
  Wlodaver, Wout, Wright, Yang, Yurdin, Zabukovic, Zachary, Zeeman, and
  Zhao]{imputation4}
Florencia Pereyra, Xiaoming Jia, Paul~J. McLaren, Amalio Telenti, Paul~I.W.
  de~Bakker, Bruce~D. Walker, Stephan Ripke, Chanson~J. Brumme, Sara~L. Pulit,
  Mary Carrington, Carl~M. Kadie, Jonathan~M. Carlson, David Heckerman,
  Robert~R. Graham, Robert~M. Plenge, Steven~G. Deeks, Lauren Gianniny, Gabriel
  Crawford, Jordan Sullivan, Elena Gonzalez, Leela Davies, Amy Camargo,
  Jamie~M. Moore, Nicole Beattie, Supriya Gupta, Andrew Crenshaw, Noël~P.
  Burtt, Candace Guiducci, Namrata Gupta, Xiaojiang Gao, Ying Qi, Yuko Yuki,
  Alicja Piechocka-Trocha, Emily Cutrell, Rachel Rosenberg, Kristin~L. Moss,
  Paul Lemay, Jessica O’leary, Todd Schaefer, Pranshu Verma, Ildiko Toth,
  Brian Block, Brett Baker, Alissa Rothchild, Jeffrey Lian, Jacqueline
  Proudfoot, Donna Marie~L. Alvino, Seanna Vine, Marylyn~M. Addo, Todd~M.
  Allen, Marcus Altfeld, Matthew~R. Henn, Sylvie~Le Gall, Hendrik Streeck,
  David~W. Haas, Daniel~R. Kuritzkes, Gregory~K. Robbins, Robert~W. Shafer,
  Roy~M. Gulick, Cecilia~M. Shikuma, Richard Haubrich, Sharon Riddler, Paul~E.
  Sax, Eric~S. Daar, Heather~J. Ribaudo, Brian Agan, Shanu Agarwal, Richard~L.
  Ahern, Brady~L. Allen, Sherly Altidor, Eric~L. Altschuler, Sujata Ambardar,
  Kathryn Anastos, Ben Anderson, Val Anderson, Ushan Andrady, Diana Antoniskis,
  David Bangsberg, Daniel Barbaro, William Barrie, J.~Bartczak, Simon Barton,
  Patricia Basden, Nesli Basgoz, Suzane Bazner, Nicholaos~C. Bellos, Anne~M.
  Benson, Judith Berger, Nicole~F. Bernard, Annette~M. Bernard, Christopher
  Birch, Stanley~J. Bodner, Robert~K. Bolan, Emilie~T. Boudreaux, Meg Bradley,
  James~F. Braun, Jon~E. Brndjar, Stephen~J. Brown, Katherine Brown, Sheldon~T.
  Brown, Jedidiah Burack, Larry~M. Bush, Virginia Cafaro, Omobolaji Campbell,
  John Campbell, Robert~H. Carlson, J.~Kevin Carmichael, Kathleen~K. Casey,
  Chris Cavacuiti, Gregory Celestin, Steven~T. Chambers, Nancy Chez, Lisa~M.
  Chirch, Paul~J. Cimoch, Daniel Cohen, Lillian~E. Cohn, Brian Conway, David~A.
  Cooper, Brian Cornelson, David~T. Cox, Michael~V. Cristofano, George
  Cuchural, Julie~L. Czartoski, Joseph~M. Dahman, Jennifer~S. Daly, Benjamin~T.
  Davis, Kristine Davis, Sheila~M. Davod, Edwin Dejesus, Craig~A. Dietz,
  Eleanor Dunham, Michael~E. Dunn, Todd~B. Ellerin, Joseph~J. Eron, John~J.W.
  Fangman, Claire~E. Farel, Helen Ferlazzo, Sarah Fidler, Anita Fleenor-Ford,
  Renee Frankel, Kenneth~A. Freedberg, Neel~K. French, Jonathan~D. Fuchs,
  Jon~D. Fuller, Jonna Gaberman, Joel~E. Gallant, Rajesh~T. Gandhi, Efrain
  Garcia, Donald Garmon, Joseph~C. Gathe, Cyril~R. Gaultier, Wondwoosen Gebre,
  Frank~D. Gilman, Ian Gilson, Paul~A. Goepfert, Michael~S. Gottlieb, Claudia
  Goulston, Richard~K. Groger, T.~Douglas Gurley, Stuart Haber, Robin
  Hardwicke, W.~David Hardy, P.~Richard Harrigan, Trevor~N. Hawkins, Sonya
  Heath, Frederick~M. Hecht, W.~Keith Henry, Melissa Hladek, Robert~P. Hoffman,
  James~M. Horton, Ricky~K. Hsu, Gregory~D. Huhn, Peter Hunt, Mark~J. Hupert,
  Mark~L. Illeman, Hans Jaeger, Robert~M. Jellinger, Mina John, Jennifer~A.
  Johnson, Kristin~L. Johnson, Heather Johnson, Kay Johnson, Jennifer Joly,
  Wilbert~C. Jordan, Carol~A. Kauffman, Homayoon Khanlou, Robert~K. Killian,
  Arthur~Y. Kim, David~D. Kim, Clifford~A. Kinder, Jeffrey~T. Kirchner, Laura
  Kogelman, Erna~Milunka Kojic, P.~Todd Korthuis, Wayne Kurisu, Douglas~S.
  Kwon, Melissa Lamar, Harry Lampiris, Massimiliano Lanzafame, Michael~M.
  Lederman, David~M. Lee, Jean~M.L. Lee, Marah~J. Lee, Edward~T.Y. Lee, Janice
  Lemoine, Jay~A. Levy, Josep~M. Llibre, Michael~A. Liguori, Susan~J. Little,
  Anne~Y. Liu, Alvaro~J. Lopez, Mono~R. Loutfy, Dawn Loy, Debbie~Y. Mohammed,
  Alan Man, Michael~K. Mansour, Vincent~C. Marconi, Martin Markowitz, Rui
  Marques, Jeffrey~N. Martin, Harold~L. Martin, Kenneth~Hugh Mayer, M.~Juliana
  McElrath, Theresa~A. McGhee, Barbara~H. McGovern, Katherine McGowan, Dawn
  McIntyre, Gavin~X. McLeod, Prema Menezes, Greg Mesa, Craig~E. Metroka, Dirk
  Meyer-Olson, Andy~O. Miller, Kate Montgomery, Karam~C. Mounzer, Ellen~H.
  Nagami, Iris Nagin, Ronald~G. Nahass, Margret~O. Nelson, Craig Nielsen,
  David~L. Norene, David~H. O’connor, Bisola~O. Ojikutu, Jason Okulicz,
  Olakunle~O. Oladehin, Edward~C. Oldfield, Susan~A. Olender, Mario Ostrowski,
  William~F. Owen, Eunice Pae, Jeffrey Parsonnet, Andrew~M. Pavlatos, Aaron~M.
  Perlmutter, Michael~N. Pierce, Jonathan~M. Pincus, Leandro Pisani,
  Lawrence~Jay Price, Laurie Proia, Richard~C. Prokesch, Heather~Calderon
  Pujet, Moti Ramgopal, Almas Rathod, Michael Rausch, J.~Ravishankar, Frank~S.
  Rhame, Constance~Shamuyarira Richards, Douglas~D. Richman, Berta Rodes,
  Milagros Rodriguez, Richard~C. Rose, Eric~S. Rosenberg, Daniel Rosenthal,
  Polly~E. Ross, David~S. Rubin, Elease Rumbaugh, Luis Saenz, Michelle~R.
  Salvaggio, William~C. Sanchez, Veeraf~M. Sanjana, Steven Santiago, Wolfgang
  Schmidt, Hanneke Schuitemaker, Philip~M. Sestak, Peter Shalit, William Shay,
  Vivian~N. Shirvani, Vanessa~I. Silebi, James~M. Sizemore, Paul~R. Skolnik,
  Marcia Sokol-Anderson, James~M. Sosman, Paul Stabile, Jack~T. Stapleton,
  Sheree Starrett, Francine Stein, Hans~Jurgen Stellbrink, F.~Lisa Sterman,
  Valerie~E. Stone, David~R. Stone, Giuseppe Tambussi, Randy~A. Taplitz,
  Ellen~M. Tedaldi, William Theisen, Richard Torres, Lorraine Tosiello, Cecile
  Tremblay, Marc~A. Tribble, Phuong~D. Trinh, Alice Tsao, Peggy Ueda, Anthony
  Vaccaro, Emilia Valadas, Thanes~J. Vanig, Isabel Vecino, Vilma~M. Vega,
  Wenoah Veikley, Barbara~H. Wade, Charles Walworth, Chingchai Wanidworanun,
  Douglas~J. Ward, Daniel~A. Warner, Robert~D. Weber, Duncan Webster, Steve
  Weis, David~A. Wheeler, David~J. White, Ed~Wilkins, Alan Winston, Clifford~G.
  Wlodaver, Angelique~Van’T Wout, David~P. Wright, Otto~O. Yang, David~L.
  Yurdin, Brandon~W. Zabukovic, Kimon~C. Zachary, Beth Zeeman, and Meng Zhao.
\newblock The major genetic determinants of hiv-1 control affect hla class i
  peptide presentation.
\newblock \emph{Science}, 330:\penalty0 1551--1557, 12 2010.
\newblock ISSN 10959203.
\newblock \doi{10.1126/science.1195271}.

\bibitem[Pillai et~al.(2014)Pillai, Okada, Saw, Ong, Wang, Tantoso, Xu,
  Peterson, Bielawny, Ali, Tay, Poh, Tan, Koo, Lim, Soong, Wenk, Raychaudhuri,
  Little, Plummer, Lee, Chia, Luo, Bakker, and Teo]{pan_asian_ref_panel1}
Nisha~Esakimuthu Pillai, Yukinori Okada, Woei~Yuh Saw, Rick Twee~Hee Ong,
  Xu~Wang, Erwin Tantoso, Wenting Xu, Trevor~A. Peterson, Thomas Bielawny,
  Mohammad Ali, Koon~Yong Tay, Wan~Ting Poh, Linda Wei~Lin Tan, Seok~Hwee Koo,
  Wei~Yen Lim, Richie Soong, Markus Wenk, Soumya Raychaudhuri, Peter Little,
  Francis~A. Plummer, Edmund~J.D. Lee, Kee~Seng Chia, Ma~Luo, Paul I.W.~De
  Bakker, and Yik~Ying Teo.
\newblock Predicting hla alleles from high-resolution snp data in three
  southeast asian populations.
\newblock \emph{Human Molecular Genetics}, 23:\penalty0 4443--4451, 2014.
\newblock ISSN 14602083.
\newblock \doi{10.1093/hmg/ddu149}.

\bibitem[Ramesh et~al.(2022)Ramesh, Dhariwal, Nichol, Chu, and Chen]{img_rcg2}
Aditya Ramesh, Prafulla Dhariwal, Alex Nichol, Casey Chu, and Mark Chen.
\newblock Hierarchical text-conditional image generation with clip latents.
\newblock 4 2022.
\newblock URL \url{http://arxiv.org/abs/2204.06125}.

\bibitem[Raychaudhuri et~al.(2012)Raychaudhuri, Sandor, Stahl, Freudenberg,
  Lee, Jia, Alfredsson, Padyukov, Klareskog, Worthington, Siminovitch, Bae,
  Plenge, Gregersen, and Bakker]{imputation5}
Soumya Raychaudhuri, Cynthia Sandor, Eli~A. Stahl, Jan Freudenberg, Hye~Soon
  Lee, Xiaoming Jia, Lars Alfredsson, Leonid Padyukov, Lars Klareskog, Jane
  Worthington, Katherine~A. Siminovitch, Sang~Cheol Bae, Robert~M. Plenge,
  Peter~K. Gregersen, and Paul I.W.~De Bakker.
\newblock Five amino acids in three hla proteins explain most of the
  association between mhc and seropositive rheumatoid arthritis.
\newblock \emph{Nature Genetics}, 44:\penalty0 291--296, 3 2012.
\newblock ISSN 10614036.
\newblock \doi{10.1038/ng.1076}.

\bibitem[Saharia et~al.(2022)Saharia, Chan, Saxena, Li, Whang, Denton,
  Ghasemipour, Ayan, Mahdavi, Lopes, Salimans, Ho, Fleet, and
  Norouzi]{img_rcg1}
Chitwan Saharia, William Chan, Saurabh Saxena, Lala Li, Jay Whang, Emily
  Denton, Seyed Kamyar~Seyed Ghasemipour, Burcu~Karagol Ayan, S.~Sara Mahdavi,
  Rapha~Gontijo Lopes, Tim Salimans, Jonathan Ho, David~J Fleet, and Mohammad
  Norouzi.
\newblock Photorealistic text-to-image diffusion models with deep language
  understanding.
\newblock 5 2022.
\newblock URL \url{http://arxiv.org/abs/2205.11487}.

\bibitem[Schuster and Paliwal(1997)]{birnn}
Mike Schuster and Kuldip~K Paliwal.
\newblock Bidirectional recurrent neural networks, 1997.

\bibitem[Todd and Beir(1987)]{dqbeta_europe1}
John~A Todd and John~I Beir.
\newblock Hla-dqp gene contributes to susceptibility and resistance to
  insulin-dependent diabetes mellitus, 1987.

\bibitem[Vaswani et~al.(2017)Vaswani, Shazeer, Parmar, Uszkoreit, Jones, Gomez,
  Kaiser, and Polosukhin]{transformer}
Ashish Vaswani, Noam Shazeer, Niki Parmar, Jakob Uszkoreit, Llion Jones,
  Aidan~N. Gomez, Lukasz Kaiser, and Illia Polosukhin.
\newblock Attention is all you need.
\newblock 6 2017.
\newblock URL \url{http://arxiv.org/abs/1706.03762}.

\bibitem[Wang et~al.(2021)Wang, Liao, Moyer, Berkowitz, Horng, and
  Golland]{ml4h_lastyear}
Peiqi Wang, Ruizhi Liao, Daniel Moyer, Seth Berkowitz, Steven Horng, and Polina
  Golland.
\newblock Image classification with consistent supporting evidence.
\newblock 11 2021.
\newblock URL \url{http://arxiv.org/abs/2111.07048}.

\bibitem[Zheng et~al.(2014)Zheng, Shen, Cox, Wakefield, Ehm, Nelson, and
  Weir]{hibag}
X.~Zheng, J.~Shen, C.~Cox, J.~C. Wakefield, M.~G. Ehm, M.~R. Nelson, and B.~S.
  Weir.
\newblock Hibag - hla genotype imputation with attribute bagging.
\newblock \emph{Pharmacogenomics Journal}, 14:\penalty0 192--200, 2014.
\newblock ISSN 14731150.
\newblock \doi{10.1038/tpj.2013.18}.

\end{thebibliography}

\appendix
\section{Imputation using bidirectional RNN}\label{apd:first}
\begin{figure}[h]
 \begin{center}
  \includegraphics[width=1.0\linewidth]{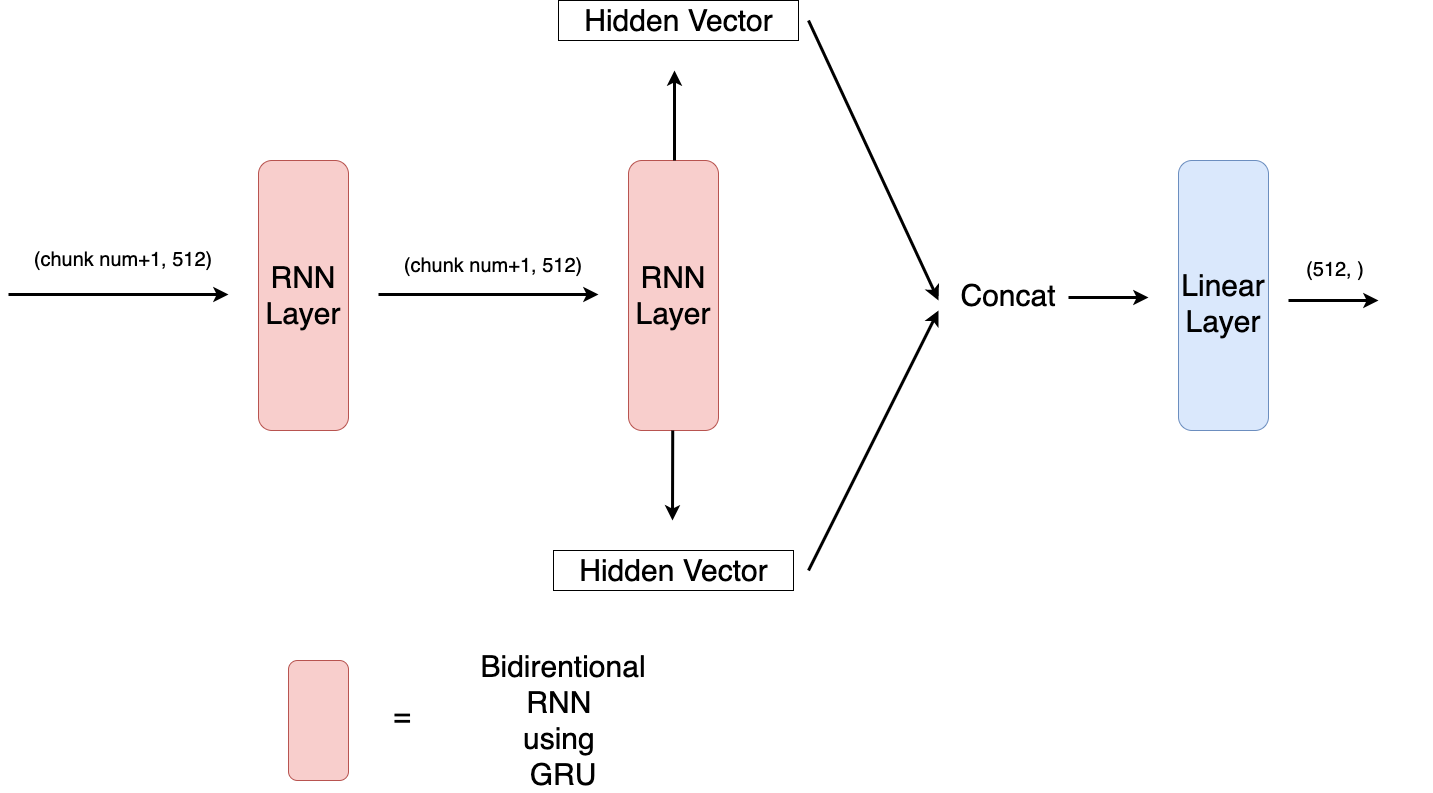}
  \caption{Architectures of the bidirectional RNN layer that we applied to "Transformer Layer" part in HLARIMNT. The hidden vectors output by the final layer of the RNN are concatenated and input to the linear layer to generate a feature vector with the dimension of 512.}
  \label{fig:fig11}
 \end{center}
\end{figure}

\begin{figure}[h]
 \begin{center}
  \includegraphics[width=0.9\linewidth]{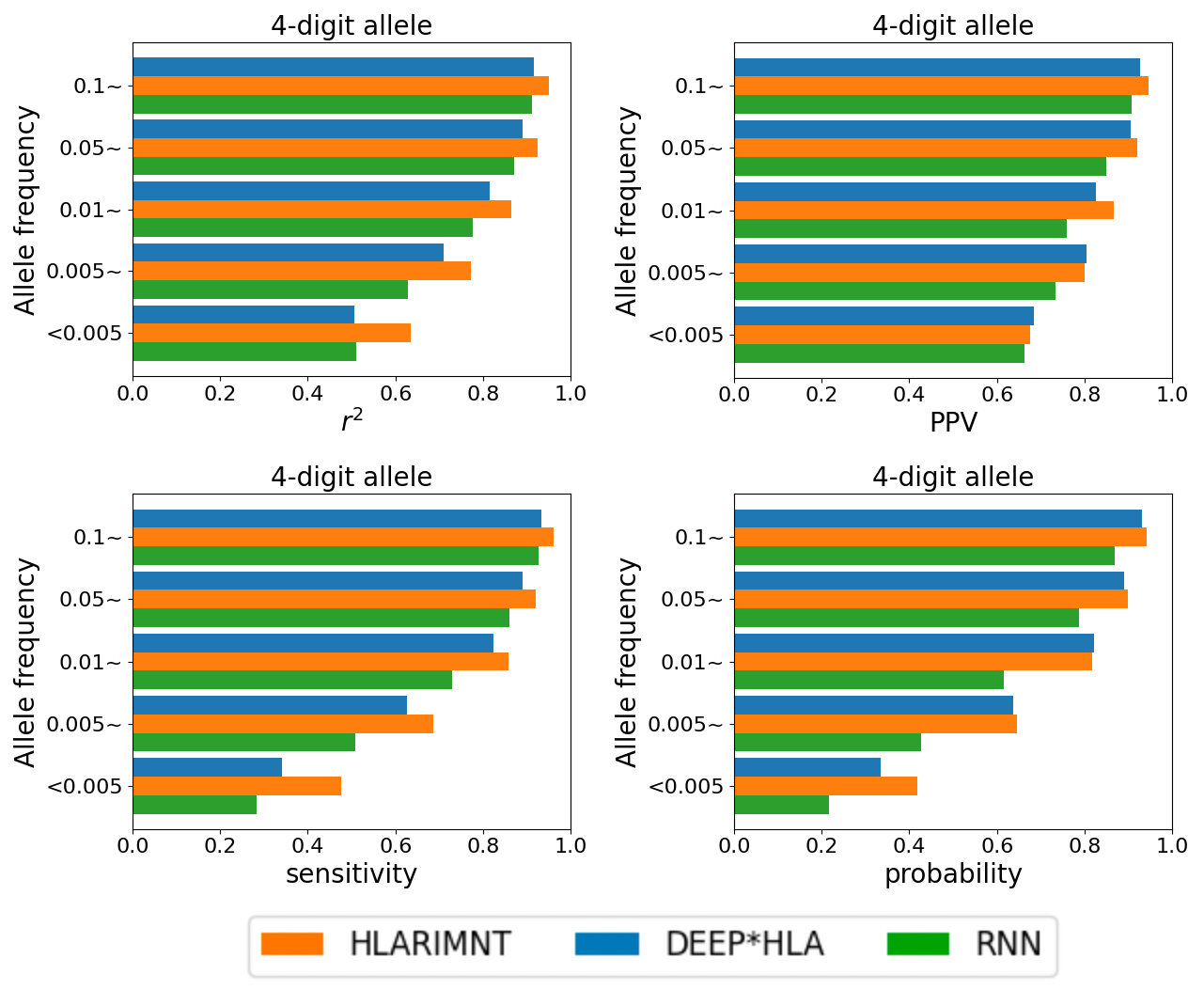}
  \caption{Accuracy of HLARIMNT, DEEP*HLA, and the bidirectional RNN. The orange bars represent the results of HLARIMNT, the blue bars represent of DEEP*HLA, and the green bars represent of the bidirectional RNN. The accuracy using the bidirectional RNN was not as good as either HLARIMNT or DEEP*HLA.}
  \label{fig:fig8}
 \end{center}
\end{figure}

We also performed an imputation using a bidirectional RNN \citep{birnn} instead of Transformer. The structure was as shown in Figure \ref{fig:fig11}; a modification of Figure \ref{fig:architectures} (c). In each of the bidirectional RNN layer, we used GRU for cells \citep{gru}. We used Optuna to search the hyperparameters. We performed Experiment 1 (\ref{subsec:exp1}) using Pan Asian reference panel, by the bidirectional RNN model. Figure \ref{fig:fig8} shows the results. The accuracy using the bidirectional RNN was not as good as either HLARIMNT or DEEP*HLA, regardless of allele frequency.

\section{Imputation using all SNPs in the reference panel}\label{apd:second}
\begin{figure}[h]
 \begin{center}
  \includegraphics[width=0.8\linewidth]{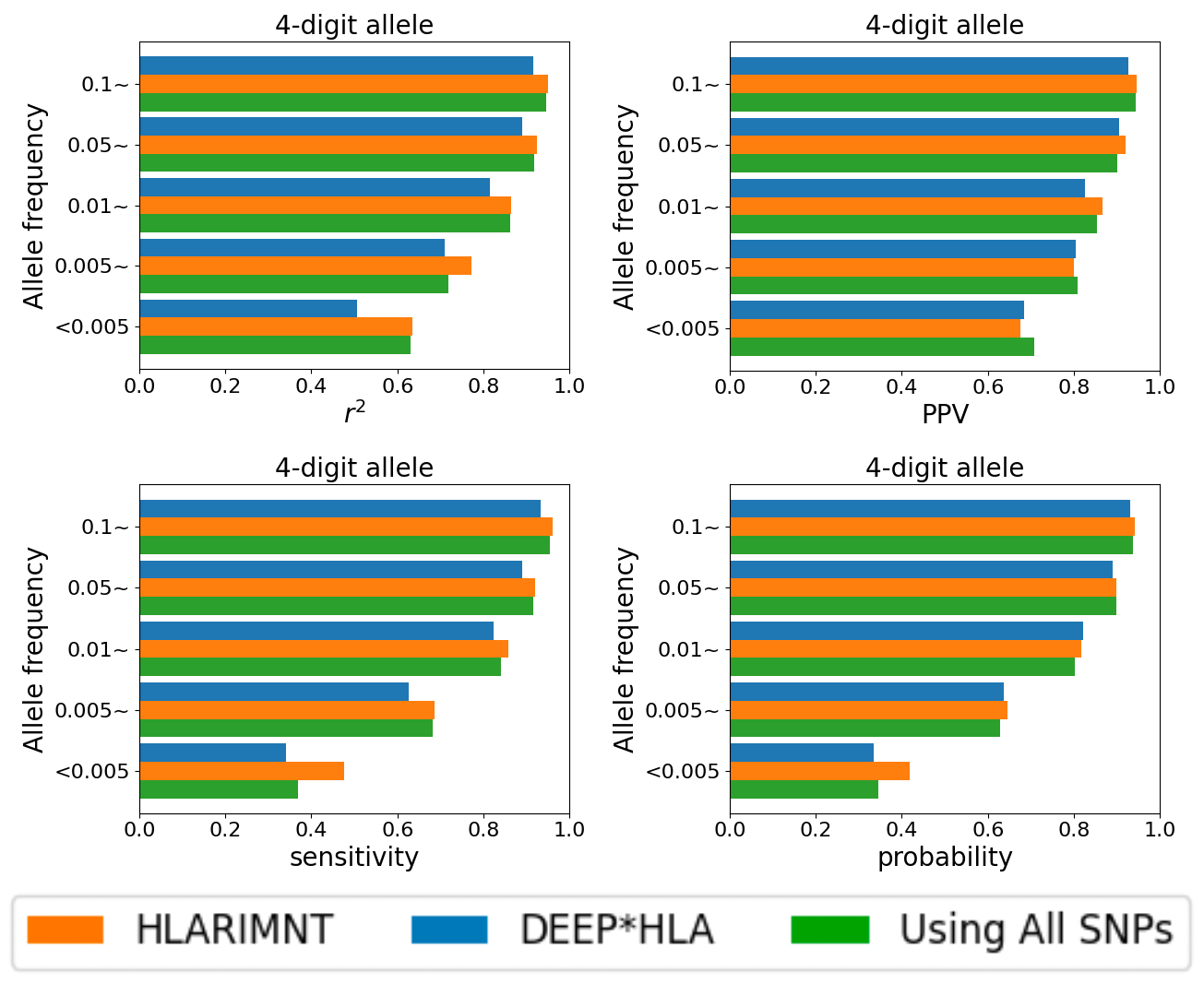}
  \caption{Accuracy of HLARIMNT using partial SNPs, DEEP*HLA, and HLARIMNT using all of the SNPs in the reference panel. The orange bars show HLARIMNT with partial SNPs as in the main article, the blue bars show DEEP*HLA, and the green bars show HLARIMNT with all SNPs. Except for PPV, HLARIMNT using partial SNPs was more accurate than using all SNPs.}
  \label{fig:fig9}
 \end{center}
\end{figure}

While the experiments in the main article used only some SNPs according to  HLA genes, we also trained our model using all of the SNPs in Pan Asian reference panel. In this case, the number of SNPs used for training was about 5 to 6 times larger, depending on the gene. The result of Experiment 1 (\ref{subsec:exp1}) under these conditions is shown in Figure \ref{fig:fig9}. HLARIMNT trained using partial SNPs were generally more accurate than that of using all SNPs, although it varied by allele frequencies. However, PPV for infrequent alleles was higher in HLARIMNT using all SNPs, the accuracy of which was even higher than that of DEEP*HLA.
\section{Cross-validation in Experiment 2}\label{apd:third}
\begin{figure}[h]
 \begin{center}
  \includegraphics[width=1.0\linewidth]{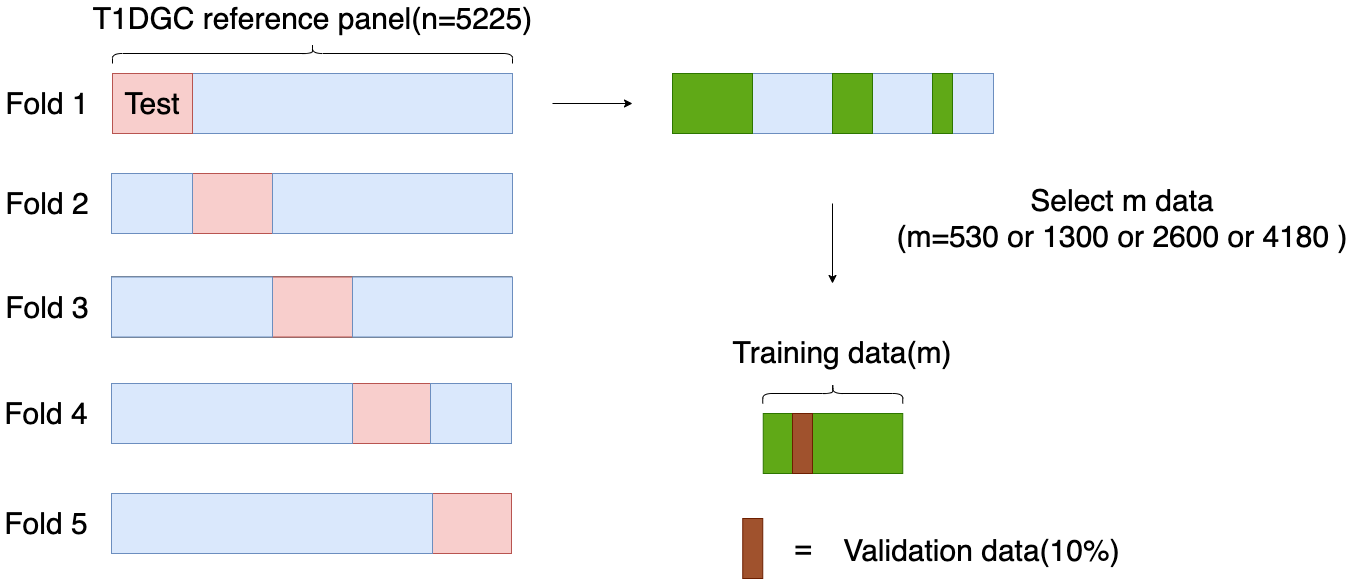}
  \caption{Details of the cross-validation in Experiment 2}
  \label{fig:fig10}
 \end{center}
\end{figure}
Here we describe the details of cross-validation  performed in Experiment 2 of the main article (\ref{subsec:exp2}). We first divided T1DGC reference panel into five parts. One partition was used as the test data for a single validation. From the remaining data, we randomly selected the data to be used for training. Here, we set the data used to train HLARIMNT and DEEP*HLA to be the same, which allowed a fair comparison of performance. We trained each model using the data selected for training, 10 percent of which was used for validation data for model updating and early stopping. 

\section{The weighted average values by allele frequency in Experiment 1}\label{apd:weighted}
\subsection{For all alleles}
In the main text we calculated a simple additive average for each allele, but here we post the weighted average values by allele frequency.
\begin{table}[h]\centering
\caption{\bfseries Average of accuracy for all 4-digit alleles weighted by allele frequency}
\label{table:table_weighted_normal}
\small
\scalebox{0.6}{
\begin{tabular}{c|c|c|c}
&&\bfseries DEEP*HLA&\bfseries HLARIMNT\\ \hline
\bfseries Pan-Asian&$\mathbf r^2$ &0.886&\bfseries0.925\\ 
&\bfseries PPV&0.898&\bfseries0.922\\ 
&\bfseries sensitivity&0.897&\bfseries0.927\\ 
&\bfseries probability&0.896&\bfseries0.903\\ \hline
\bfseries T1DGC&$\mathbf r^2$ &0.971& \bfseries0.976 \\ 
&\bfseries PPV&\bfseries0.976&0.975\\ 
&\bfseries sensitivity&0.974&\bfseries0.977\\ 
&\bfseries probability&\bfseries0.974&\bfseries0.974\\ \hline
\bfseries Mixed&$\mathbf r^2$ & 0.918&\bfseries 0.935\\ 
&\bfseries PPV&0.927&\bfseries0.935\\ 
&\bfseries sensitivity&0.928&\bfseries0.936\\ 
&\bfseries probability&\bfseries0.927&0.926\\ \hline
\end{tabular}
}
\end{table}

\subsection{For infrequent alleles}
\begin{table}[h]\centering
\caption{\bfseries Average of accuracy for infrequent 4-digit alleles weighted by allele frequency}
\label{table:table_weighted_rare}
\small
\scalebox{0.6}{
\begin{tabular}{c|c|c|c}
&&\bfseries DEEP*HLA&\bfseries HLARIMNT\\ \hline
\bfseries Pan-Asian&$\mathbf r^2$ &0.664&\bfseries0.740\\ 
&\bfseries PPV&\bfseries0.781&0.759\\ 
&\bfseries sensitivity&0.550&\bfseries0.641\\ 
&\bfseries probability&0.554&\bfseries0.586\\ \hline
\bfseries T1DGC&$\mathbf r^2$ &0.852& \bfseries0.866 \\ 
&\bfseries PPV&\bfseries0.882&0.875\\ 
&\bfseries sensitivity&0.840&\bfseries0.842\\ 
&\bfseries probability&\bfseries0.838&0.831\\ \hline
\bfseries Mixed&$\mathbf r^2$ & 0.736&\bfseries 0.777\\ 
&\bfseries PPV&0.784&\bfseries0.794\\ 
&\bfseries sensitivity&0.698&\bfseries0.733\\ 
&\bfseries probability&0.696&\bfseries0.706\\ \hline
\end{tabular}
}
\end{table}

\section{Confidence intervals for Experiment 1}\label{apd:interval_1}

Here we report 95\% confidence intervals for each test data used in Experiment 1. The very wide confidence intervals are due to the fact that the accuracy varies extremely with the allele frequency, which makes the variance very large.

\begin{table}[h]\centering
\caption{\bfseries Confidence intervals for 4-digit alleles in test data (1)}
\label{table:interval_1}
\small
\scalebox{0.6}{
\begin{tabular}{c|c|c|c}
&&\bfseries DEEP*HLA&\bfseries HLARIMNT\\ \hline
\bfseries Pan-Asian&$\mathbf r^2$ &$0.814\pm0.096$&$\mathbf{0.876\pm0.033}$\\ 
&\bfseries PPV&$0.845\pm0.069$&$\mathbf{0.88\pm0.035}$\\ 
&\bfseries sensitivity&$0.785\pm0.094$&$\mathbf{0.832\pm0.047}$\\ 
&\bfseries probability&$0.787\pm0.054$&$\mathbf{0.797\pm0.044}$\\ \hline
\bfseries T1DGC&$\mathbf r^2$ &$0.831\pm0.057$& $\mathbf{0.857\pm0.032}$ \\ 
&\bfseries PPV&$\mathbf{0.897\pm0.002}$&$0.866\pm0.033$\\ 
&\bfseries sensitivity&$0.783\pm0.081$&$\mathbf{0.825\pm0.039}$\\ 
&\bfseries probability&$0.784\pm0.058$&\bfseries$\mathbf{0.805\pm0.038}$\\ \hline
\bfseries Mixed&$\mathbf r^2$ & $0.794\pm0.084$&$\mathbf{0.841\pm0.037}$\\ 
&\bfseries PPV&$0.845\pm0.051$&$\mathbf{0.861\pm0.036}$\\ 
&\bfseries sensitivity&$0.745\pm0.097$&$\mathbf{0.794\pm0.048}$\\ 
&\bfseries probability&$0.746\pm0.079$&$\mathbf{0.780\pm0.044}$\\ \hline
\end{tabular}
}
\end{table}

\begin{table}[h]\centering
\caption{\bfseries Confidence intervals for 4-digit alleles in test data (2)}

\small
\scalebox{0.6}{
\begin{tabular}{c|c|c|c}
&&\bfseries DEEP*HLA&\bfseries HLARIMNT\\ \hline
\bfseries Pan-Asian&$\mathbf r^2$ &$0.823\pm0.087$&$\mathbf{0.880\pm0.030}$\\ 
&\bfseries PPV&$0.840\pm0.071$&$\mathbf{0.876\pm0.036}$\\ 
&\bfseries sensitivity&$0.792\pm0.103$&$\mathbf{0.854\pm0.040}$\\ 
&\bfseries probability&$0.790\pm0.066$&$\mathbf{0.818\pm0.038}$\\ \hline
\bfseries T1DGC&$\mathbf r^2$ &$0.812\pm0.052$& $\mathbf{0.830\pm0.035}$ \\ 
&\bfseries PPV&$\mathbf{0.886\pm0.019}$&$0.873\pm0.032$\\ 
&\bfseries sensitivity&$0.786\pm0.062$&$\mathbf{0.808\pm0.041}$\\ 
&\bfseries probability&$0.787\pm0.047$&\bfseries$\mathbf{0.795\pm0.039}$\\ \hline
\bfseries Mixed&$\mathbf r^2$ & $0.828\pm0.068$&$\mathbf{0.866\pm0.030}$\\ 
&\bfseries PPV&$0.867\pm0.041$&$\mathbf{0.876\pm0.031}$\\ 
&\bfseries sensitivity&$0.788\pm0.093$&$\mathbf{0.843\pm0.039}$\\ 
&\bfseries probability&$0.788\pm0.068$&$\mathbf{0.819\pm0.038}$\\ \hline
\end{tabular}
}
\end{table}

\begin{table}[h]\centering
\caption{\bfseries Confidence intervals for 4-digit alleles in test data (3)}

\small
\scalebox{0.6}{
\begin{tabular}{c|c|c|c}
&&\bfseries DEEP*HLA&\bfseries HLARIMNT\\ \hline
\bfseries Pan-Asian&$\mathbf r^2$ &$0.776\pm0.096$&$\mathbf{0.832\pm0.040}$\\ 
&\bfseries PPV&$0.849\pm0.042$&$\mathbf{0.850\pm0.041}$\\ 
&\bfseries sensitivity&$0.775\pm0.079$&$\mathbf{0.805\pm0.050}$\\ 
&\bfseries probability&$\mathbf{0.771\pm0.038}$&$0.761\pm0.048$\\ \hline
\bfseries T1DGC&$\mathbf r^2$ &$0.818\pm0.056$& $\mathbf{0.840\pm0.034}$ \\ 
&\bfseries PPV&$0.896\pm0.033$&$\mathbf{0.904\pm0.025}$\\ 
&\bfseries sensitivity&$0.774\pm0.062$&$\mathbf{0.793\pm0.042}$\\ 
&\bfseries probability&$0.775\pm0.056$&\bfseries$\mathbf{0.791\pm0.041}$\\ \hline
\bfseries Mixed&$\mathbf r^2$ & $0.809\pm0.060$&$\mathbf{0.830\pm0.039}$\\ 
&\bfseries PPV&$\mathbf{0.893\pm0.019}$&$0.879\pm0.033$\\ 
&\bfseries sensitivity&$0.766\pm0.090$&$\mathbf{0.814\pm0.042}$\\ 
&\bfseries probability&$0.767\pm0.078$&$\mathbf{0.803\pm0.042}$\\ \hline
\end{tabular}
}
\end{table}

\begin{table}[h]\centering
\caption{\bfseries Confidence intervals for 4-digit alleles in test data (4)}

\small
\scalebox{0.6}{
\begin{tabular}{c|c|c|c}
&&\bfseries DEEP*HLA&\bfseries HLARIMNT\\ \hline
\bfseries Pan-Asian&$\mathbf r^2$ &$0.795\pm0.088$&$\mathbf{0.847\pm0.037}$\\ 
&\bfseries PPV&$0.834\pm0.063$&$\mathbf{0.860\pm0.037}$\\ 
&\bfseries sensitivity&$0.775\pm0.089$&$\mathbf{0.817\pm0.047}$\\ 
&\bfseries probability&$0.771\pm0.050$&$\mathbf{0.776\pm0.045}$\\ \hline
\bfseries T1DGC&$\mathbf r^2$ &$0.812\pm0.044$& $\mathbf{0.820\pm0.036}$ \\ 
&\bfseries PPV&$\mathbf{0.898\pm0.005}$&$0.870\pm0.033$\\ 
&\bfseries sensitivity&$\mathbf{0.778\pm0.040}$&$0.772\pm0.045$\\ 
&\bfseries probability&$\mathbf{0.776\pm0.035}$&\bfseries$0.768\pm0.043$\\ \hline
\bfseries Mixed&$\mathbf r^2$ & $0.782\pm0.084$&$\mathbf{0.828\pm0.039}$\\ 
&\bfseries PPV&$0.850\pm0.042$&$\mathbf{0.857\pm0.035}$\\ 
&\bfseries sensitivity&$0.752\pm0.084$&$\mathbf{0.790\pm0.047}$\\ 
&\bfseries probability&$0.752\pm0.069$&$\mathbf{0.777\pm0.044}$\\ \hline
\end{tabular}
}
\end{table}

\begin{table}[h]\centering
\caption{\bfseries Confidence intervals for 4-digit alleles in test data (5)}

\small
\scalebox{0.6}{
\begin{tabular}{c|c|c|c}
&&\bfseries DEEP*HLA&\bfseries HLARIMNT\\ \hline
\bfseries Pan-Asian&$\mathbf r^2$ &$0.767\pm0.092$&$\mathbf{0.815\pm0.043}$\\ 
&\bfseries PPV&$0.851\pm0.049$&$\mathbf{0.859\pm0.041}$\\ 
&\bfseries sensitivity&$0.739\pm0.109$&$\mathbf{0.793\pm0.054}$\\ 
&\bfseries probability&$0.736\pm0.083$&$\mathbf{0.768\pm0.051}$\\ \hline
\bfseries T1DGC&$\mathbf r^2$ &$0.840\pm0.043$& $\mathbf{0.848\pm0.035}$ \\ 
&\bfseries PPV&$\mathbf{0.910\pm0.011}$&$0.890\pm0.031$\\ 
&\bfseries sensitivity&$0.811\pm0.049$&$\mathbf{0.820\pm0.039}$\\ 
&\bfseries probability&$0.810\pm0.042$&\bfseries$\mathbf{0.813\pm0.038}$\\ \hline
\bfseries Mixed&$\mathbf r^2$ & $0.763\pm0.064$&$\mathbf{0.785\pm0.042}$\\ 
&\bfseries PPV&$0.855\pm0.052$&$\mathbf{0.875\pm0.032}$\\ 
&\bfseries sensitivity&$\mathbf{0.755\pm0.041}$&$0.746\pm0.050$\\ 
&\bfseries probability&$\mathbf{0.750\pm0.010}$&$0.712\pm0.048$\\ \hline
\end{tabular}
}
\end{table}
\section{Confidence intervals of infrequent alleles for Experiment 1}\label{apd:interval_2}

Here we report 95\% confidence intervals of accuracy for infrequent alleles, on each test data used in Experiment 1. The very wide confidence intervals are due to the fact that the accuracy varies extremely with the allele frequency, which makes the variance very large. Furthermore, the limited number of infrequent alleles is another reason for the wide confidence intervals.

\begin{table}[h]\centering
\caption{\bfseries Confidence intervals for infrequent 4-digit alleles in test data (1)}
\label{table:interval_2}
\small
\scalebox{0.6}{
\begin{tabular}{c|c|c|c}
&&\bfseries DEEP*HLA&\bfseries HLARIMNT\\ \hline
\bfseries Pan-Asian&$\mathbf r^2$ &$0.595\pm0.164$ &$\mathbf{0.757\pm0.130}$\\ 
&\bfseries PPV&\bfseries$0.697\pm0.195$&$\mathbf{0.803\pm0.151}$\\ 
&\bfseries sensitivity&$0.430\pm0.169$&$\mathbf{0.548\pm0.167}$\\ 
&\bfseries probability&$0.447\pm0.165$&$\mathbf{0.498\pm0.137}$\\ \hline
\bfseries T1DGC&$\mathbf r^2$ &$0.721\pm0.057$ &$\mathbf{0.765\pm0.052}$\\ 
&\bfseries PPV&$\mathbf{0.829\pm0.052}$&$0.777\pm0.058$\\ 
&\bfseries sensitivity&$0.631\pm0.068$&$\mathbf{0.709\pm0.064}$\\ 
&\bfseries probability&$0.633\pm0.068$&$\mathbf{0.679\pm0.060}$\\ \hline
\bfseries Mixed&$\mathbf r^2$ &$0.641\pm0.086$&$\mathbf{0.713\pm0.080}$\\ 
&\bfseries PPV&$0.725\pm0.096$&$\mathbf{0.748\pm0.088}$\\ 
&\bfseries sensitivity&$0.509\pm0.100$&$\mathbf{0.601\pm0.098}$\\ 
&\bfseries probability&$0.512\pm0.098$&$\mathbf{0.594\pm0.089}$\\ \hline

\end{tabular}
}
\end{table}
\begin{table}[h]\centering
\caption{\bfseries Confidence intervals for infrequent 4-digit alleles in test data (2)}

\small
\scalebox{0.6}{
\begin{tabular}{c|c|c|c}
&&\bfseries DEEP*HLA&\bfseries HLARIMNT\\ \hline
\bfseries Pan-Asian&$\mathbf r^2$ &$0.678\pm0.144$&$\mathbf{0.811\pm0.122}$\\ 
&\bfseries PPV&$0.682\pm0.177$&$\mathbf{0.786\pm0.146}$\\ 
&\bfseries sensitivity&$0.511\pm0.173$&$\mathbf{0.742\pm0.157}$\\ 
&\bfseries probability&$0.512\pm0.166$&$\mathbf{0.621\pm0.129}$\\ \hline
\bfseries T1DGC&$\mathbf r^2$ &$0.701\pm0.059$& $\mathbf{0.725\pm0.055}$ \\ 
&\bfseries PPV&$\mathbf{0.814\pm0.049}$&$0.794\pm0.055$\\ 
&\bfseries sensitivity&$0.656\pm0.067$&$\mathbf{0.688\pm0.064}$\\ 
&\bfseries probability&$0.658\pm0.066$&\bfseries$\mathbf{0.669\pm0.061}$\\ \hline
\bfseries Mixed&$\mathbf r^2$ & $0.692\pm0.083$&$\mathbf{0.758\pm0.074}$\\ 
&\bfseries PPV&$0.764\pm0.091$&$\mathbf{0.791\pm0.082}$\\ 
&\bfseries sensitivity&$0.573\pm0.106$&$\mathbf{0.686\pm0.094}$\\ 
&\bfseries probability&$0.573\pm0.105$&$\mathbf{0.649\pm0.086}$\\ \hline
\end{tabular}
}
\end{table}

\begin{table}[h]\centering
\caption{\bfseries Confidence intervals for infrequent 4-digit alleles in test data (3)}

\small
\scalebox{0.6}{
\begin{tabular}{c|c|c|c}
&&\bfseries DEEP*HLA&\bfseries HLARIMNT\\ \hline
\bfseries Pan-Asian&$\mathbf r^2$ &$0.543\pm0.162$&$\mathbf{0.623\pm0.141}$\\ 
&\bfseries PPV&$\mathbf{0.808\pm0.163}$&$0.660\pm0.177$\\ 
&\bfseries sensitivity&$0.473\pm0.165$&$\mathbf{0.492\pm0.162}$\\ 
&\bfseries probability&$\mathbf{0.466\pm0.162}$&$0.460\pm0.144$\\ \hline
\bfseries T1DGC&$\mathbf r^2$ &$0.714\pm0.059$& $\mathbf{0.747\pm0.053}$ \\ 
&\bfseries PPV&$0.838\pm0.049$&$\mathbf{0.849\pm0.045}$\\ 
&\bfseries sensitivity&$0.636\pm0.069$&$\mathbf{0.667\pm0.066}$\\ 
&\bfseries probability&$0.638\pm0.068$&\bfseries$\mathbf{0.664\pm0.063}$\\ \hline
\bfseries Mixed&$\mathbf r^2$ & $0.675\pm0.084$&$\mathbf{0.689\pm0.085}$\\ 
&\bfseries PPV&$\mathbf{0.870\pm0.064}$&$0.802\pm0.081$\\ 
&\bfseries sensitivity&$0.555\pm0.097$&$\mathbf{0.656\pm0.089}$\\ 
&\bfseries probability&$0.559\pm0.096$&$\mathbf{0.639\pm0.087}$\\ \hline
\end{tabular}
}
\end{table}

\begin{table}[h]\centering
\caption{\bfseries Confidence intervals for infrequent 4-digit alleles in test data (4)}

\small
\scalebox{0.6}{
\begin{tabular}{c|c|c|c}
&&\bfseries DEEP*HLA&\bfseries HLARIMNT\\ \hline
\bfseries Pan-Asian&$\mathbf r^2$ &$0.591\pm0.158$&$\mathbf{0.665\pm0.138}$\\ 
&\bfseries PPV&$\mathbf{0.714\pm0.182}$&$0.663\pm0.171$\\ 
&\bfseries sensitivity&$0.483\pm0.174$&$\mathbf{0.529\pm0.174}$\\ 
&\bfseries probability&$0.476\pm0.172$&$\mathbf{0.486\pm0.148}$\\ \hline
\bfseries T1DGC&$\mathbf r^2$ &$0.698\pm0.061$& $\mathbf{0.711\pm0.056}$ \\ 
&\bfseries PPV&$\mathbf{0.829\pm0.051}$&$0.785\pm0.059$\\ 
&\bfseries sensitivity&$\mathbf{0.636\pm0.069}$&$0.628\pm0.069$\\ 
&\bfseries probability&$\mathbf{0.634\pm0.069}$&\bfseries$0.624\pm0.065$\\ \hline
\bfseries Mixed&$\mathbf r^2$ & $0.589\pm0.095$&$\mathbf{0.667\pm0.087}$\\ 
&\bfseries PPV&$\mathbf{0.727\pm0.104}$&$\mathbf{0.727\pm0.095}$\\ 
&\bfseries sensitivity&$0.505\pm0.104$&$\mathbf{0.572\pm0.102}$\\ 
&\bfseries probability&$0.505\pm0.103$&$\mathbf{0.559\pm0.092}$\\ \hline
\end{tabular}
}
\end{table}

\begin{table}[h]\centering
\caption{\bfseries Confidence intervals for infrequent 4-digit alleles in test data (5)}

\small
\scalebox{0.6}{
\begin{tabular}{c|c|c|c}
&&\bfseries DEEP*HLA&\bfseries HLARIMNT\\ \hline
\bfseries Pan-Asian&$\mathbf r^2$ &$0.506\pm0.152$&$\mathbf{0.569\pm0.135}$\\ 
&\bfseries PPV&$\mathbf{0.800\pm0.186}$&$0.712\pm0.165$\\ 
&\bfseries sensitivity&$0.338\pm0.152$&$\mathbf{0.454\pm0.158}$\\ 
&\bfseries probability&$0.330\pm0.146$&$\mathbf{0.444\pm0.145}$\\ \hline
\bfseries T1DGC&$\mathbf r^2$ &$0.741\pm0.058$& $\mathbf{0.750\pm0.057}$ \\ 
&\bfseries PPV&$\mathbf{0.862\pm0.047}$&$0.821\pm0.054$\\ 
&\bfseries sensitivity&$0.686\pm0.067$&$\mathbf{0.700\pm0.064}$\\ 
&\bfseries probability&$0.683\pm0.067$&\bfseries$\mathbf{0.689\pm0.061}$\\ \hline
\bfseries Mixed&$\mathbf r^2$ & $0.544\pm0.092$&$\mathbf{0.578\pm0.083}$\\ 
&\bfseries PPV&$0.727\pm0.096$&$\mathbf{0.775\pm0.086}$\\ 
&\bfseries sensitivity&$\mathbf{0.513\pm0.100}$&$0.488\pm0.096$\\ 
&\bfseries probability&$\mathbf{0.504\pm0.098}$&$0.442\pm0.084$\\ \hline
\end{tabular}
}
\end{table}

\section{Training Flow}\label{apd:train_flow}
First, a variant 'count', which is an indicator used for early stopping, is initialized to 0. At the beginning of the epoch, the training data is batched and trained using back propagation. At the end of one epoch, the model's correctness rate (the percentage of data that the model could impute correctly) in the validation data is calculated. If this value is greater than the value of the ‘best model’, the weights are overwritten and saved as the ‘best model’. If not, add 1 to the ‘count’, and when the value of the ‘count’ reaches specified number, the training is terminated. Also, at regular intervals, the learning rate is decreased. This process is repeated until the number of epochs exceeds specified number, at which point training is terminated. The ‘best model’ stored at the end of training is used to examine the accuracy for the test data.

\section{Experiment1 additional figures}
\label{apd:exp1}
\begin{figure}[h]
 \begin{center}
  \includegraphics[width=0.69\linewidth]{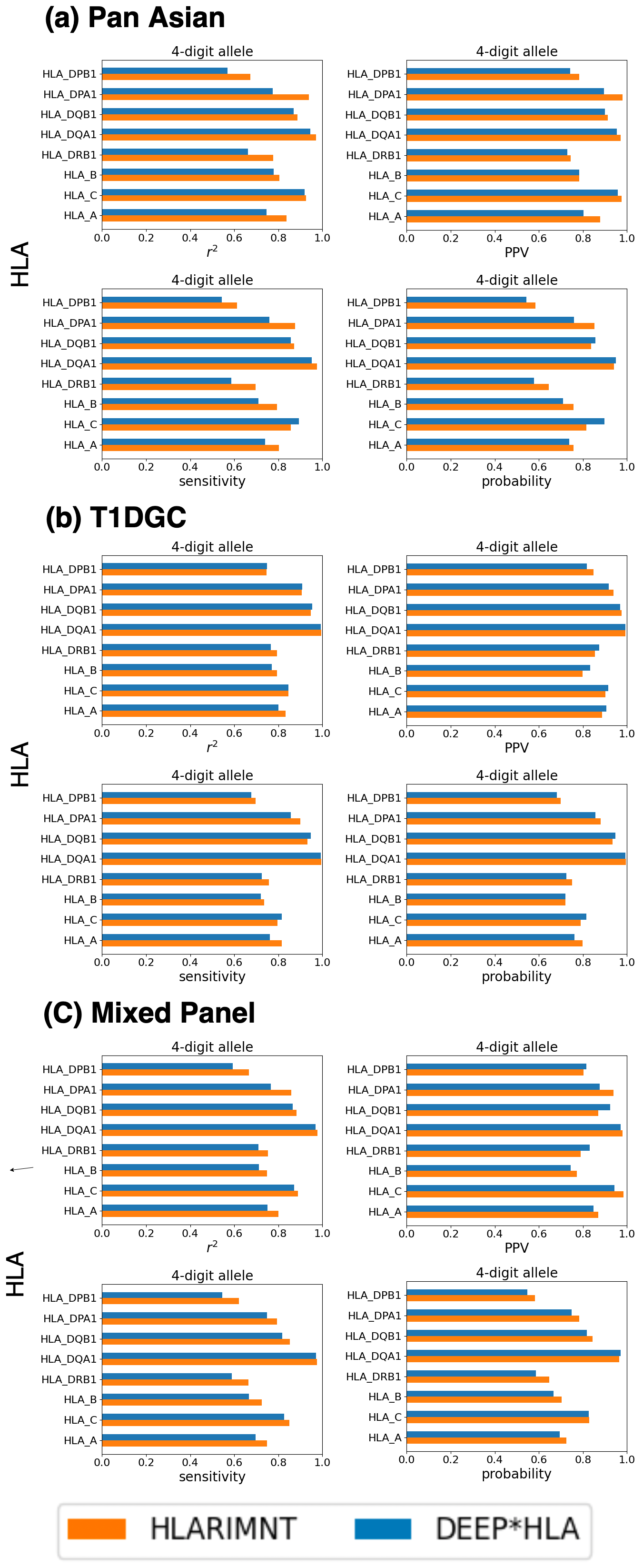}
  \caption{Average accuracy for 4-digit alleles in each of the HLA genes. HLARIMNT outperformed DEEP*HLA on most genes and indices for all datasets, though there were some exceptions depending on the combination of indices and genes.}
  \label{fig:fig3}
 \end{center}
\end{figure}

Average imputation accuracy for 4-digit alleles in each HLA gene is showed in Figure \ref{fig:fig3}. HLARIMNT generally showed advantages for almost all of the genes in all of the three reference panels, with some exceptions depending on the combination of genes, reference panels, and indices.
\begin{figure}[h]
 \begin{center}
  \includegraphics[width=0.69\linewidth]{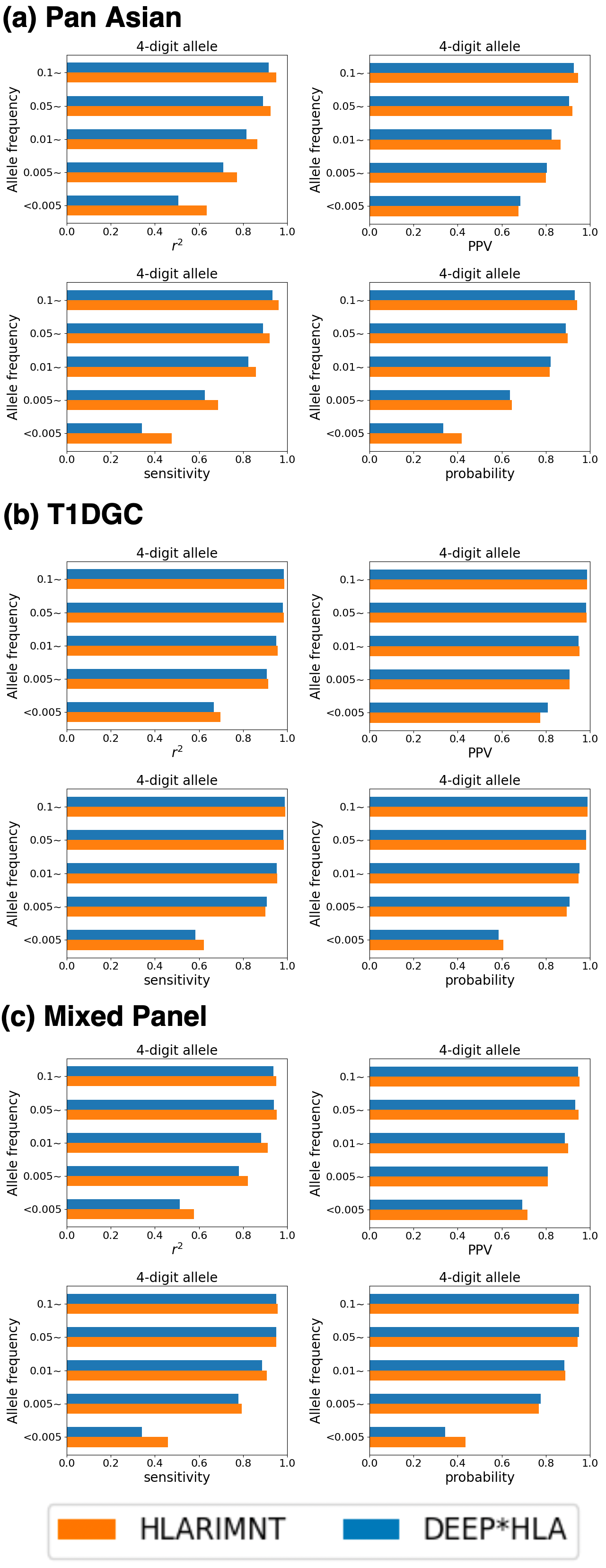}
  \caption{Average values of the 4 indices calculated for each allele frequency. At almost all of the frequencies, the accuracy of HLARIMNT was equal to or better than DEEP*HLA. Furthermore, the advantage of HLARIMNT was more noticeable at lower frequencies.}
  \label{fig:fig4}
 \end{center}
\end{figure}
In Figure \ref{fig:fig4}, for each metric, accuracy for 4-digit alleles with frequencies equal to or above the value on the vertical axis is shown on the horizontal axis (Notice that $<0.005$ indicates a frequency of less than 0.005). 
When using Pan-Asian reference panel and the mixed counterpart, HLARIMNT performed with almost the same or better accuracy than DEEP*HLA regardless of allele frequencies. The superiority of HLARIMNT was more noticeable for alleles with lower frequencies. When using T1DGC reference panel, though there was no marked difference for frequent alleles, the accuracy for extremely rare alleles was apparently higher in HLARIMNT except for PPV.
\section{Hyperparameters}
\label{apd:hyper}
 We used Optuna (\url{https://github.com/optuna/optuna}) to search hyperparameters. The searched parameters and the values are as follows;
Number of Transformer heads: 64,
Number of layers of Transformer encoder: 2,
Dimension of feedforward in Transformer encoder: 64,
Batch size: 64,
Dimension of embedding in Embedding Layer: 512, Learning rate:  0.0005, Number of epochs for early stopping: 50.
\section{Further Discussion}
\label{apd:discussion}
Here is a more detailed discussion of the study and future perspectives.\\
\indent What is noteworthy about Experiment 1 is that HLARIMNT outperformed in a mixed panel of two ethnic groups. This indicates that HLARIMNT captured features of the SNPs without over-fitting a particular ethnic group. This characteristic may be useful in practical HLA imputation; For an accurate HLA imputation, a large reference panel is required. In fact, in Experiment 2, both methods were more accurate with a larger number of training data. On the other hand, it is known that the distribution and and frequency of the HLA alleles are variable across different ethnic groups \citep{distri_by_eth}, which results in heterogeneity in HLA risk alleles across populations \citep{pan_asian_ref_panel2}. This phenomenon is seen, for example, in the association between non-Asp57 in HLA-DQB1 and type1 diabetes (T1D) risk; in Europeans, there is a strong correlation between these two \citep{dqbeta_europe1,dqbeta_europe2}, but not in Japanese \citep{dqbeta_jap}. Therefore, it is desirable to create large reference panels for each race. However, to create a large reference panel, it is necessary to analyze SNPs of many individuals at high density, which is very expensive. In this regard, if we can use the mixture of reference panels including various ethnic groups for training, it will be possible to perform an accurate HLA imputation without performing new SNP sequencing to make the reference panel larger. HLARIMNT has the potential to meet this requirement.\\
\indent What should be mentioned about Experiment 2 is that the superiority of HLARIMNT was more pronounced with less training data, regardless of allele frequencies. Basically, while haplotypes vary by ethnicity, a reference panel needs to be created for each race. However, it is expensive to create a new reference panel for a specific race. HLARIMNT could solve this problem, by capturing the characteristics  with less data more accurately.\\
\indent For these reasons, HLARIMNT is expected to become a practical deep learning method for HLA imputation. As mentioned in section \ref{sec:intro}, research over the past few years has shown that Transformer is useful for sequential data. In addition, this study strongly suggests that Transformer may be more useful than conventional methods, like RNN (Appendix \ref{apd:first}) as well as CNN, in the modality of genomic information. This should be an opportunity to experiment more with Transformer in the field of genomics.\\
\indent However, there are several points that should be discussed in this study. One is the small number of individuals (530 individuals; 1060 haplotypes) in Pan-Asian reference panel. In this case, alleles with frequencies of 0.005 or low are very few, thus accuracy for them may be unreliable, especially in indices such as probability. In addition, in both experiments, there were some situations in which HLARIMNT lacked pronounced priority or was inferior in PPV, especially for infrequent alleles. The combination of high sensitivity and low PPV in infrequent alleles means that HLARIMNT is finding characteristics of infrequent alleles in many other alleles. A more detailed analysis of this error may reveal how HLARIMNT is making the allele distinction, and it may allow us to make more accurate imputation models. Finally, while the advantage of HLARIMNT has already been amply demonstrated, it may be possible to evaluate its utility further by making test data of different ethnicity from training data, or by seeing if there is a difference in accuracy among each SNPs sequence. In this case, ethnic groups that are closely related will be selected for training and test data for appropriate validation, as there is a large difference in alleles among ethnicities as mentioned above.

\end{document}